%% file: main.tex
\pdfoutput=1
\documentclass[11pt, a4paper]{article}

\usepackage[utf8]{inputenc}
\usepackage[T1]{fontenc}
\usepackage{lmodern}
\usepackage[margin=1in]{geometry}

\usepackage{authblk}
\usepackage{orcidlink}

\usepackage{hyperref}
\usepackage[
    backend=biber,
    style=authoryear,       
    natbib=true,        
    giveninits=true,  
    uniquename=init,
    doi=true,           
    url=true,
    maxcitenames=2,
    mincitenames=1,
    maxbibnames=99,
    uniquelist=false,
    isbn=false
]{biblatex}
\DeclareNameAlias{author}{family-given}
\DeclareFieldFormat{pages}{#1}

\AtEveryBibitem{\clearfield{note}}
\AtEveryBibitem{\clearfield{urldate}}
\AtEveryBibitem{\clearlist{language}}
\AtEveryBibitem{\clearlist{location}}

\DeclareFieldFormat*{title}{#1}

\renewbibmacro{in:}{%
  {}{\printtext{\bibstring{in}\intitlepunct}}%
}

\AtEveryBibitem{%
  \ifboolexpr{
    not test {}
    or
    not test {}
  }
    {\clearfield{url}}
    {}%
}

\usepackage{mathbbol} %
\usepackage{amsmath, amsthm, amssymb, amsfonts, bm}

\DeclareSymbolFontAlphabet{\mathbb}{AMSb}
\DeclareSymbolFontAlphabet{\mathbbl}{bbold}  

\usepackage{mathtools}
\mathtoolsset{showonlyrefs=true}
\usepackage{bbm}
\usepackage{siunitx}
\usepackage{hyperref}
\usepackage[thinlines]{easytable}

\usepackage{enumitem}
\usepackage{graphicx}
\usepackage{comment}
\usepackage{aligned-overset}
\usepackage[dvipsnames]{xcolor}
\usepackage{soul}
\usepackage{subfig}
\definecolor{lightgray}{gray}{0.8}

\usepackage{booktabs} 
\usepackage{multirow} 

\usepackage{algorithm}
\usepackage{algpseudocode}

\usepackage{tikz}
\usetikzlibrary{positioning, shapes, patterns, arrows.meta, calc}

\newtheorem{theorem}{Theorem}[section]
\newtheorem{corollary}{Corollary}[section]
\newtheorem{lemma}{Lemma}[section]
\newtheorem{proposition}{Proposition}[section]
\newtheorem{example}{Example}[section]
\theoremstyle{definition}
\newtheorem{definition}{Definition}[section]

\newtheorem{remark}{Remark}[section]

\newtheorem{assumption}{Assumption}

\newcommand{\norm}[1]{\lVert#1\rVert}
\newcommand{\snorm}[1]{{\left\vert\kern-0.25ex\left\vert\kern-0.25ex\left\vert #1 
    \right\vert\kern-0.25ex\right\vert\kern-0.25ex\right\vert}}
\newcommand{\abs}[1]{\left|#1\right|}
\newcommand{\expv}[1]{\mathbb{E} \left[#1\right]}
\newcommand{\prob}[1]{\mathbb{P}\left(#1\right)}
\newcommand{\cov}{\mathbb{C}\textup{ov}}
\newcommand{\var}{\mathbb{V}}

\newcommand{\ind}{\mathbbm{1}}

\newcommand{\tr}[1]{\text{tr}\left(#1\right)}

\renewbibmacro{in:}{%
  \ifentrytype{article}{}{\printtext{\bibstring{in}\intitlepunct}}%
}

\newcommand{\dist}[0]{\text{dist}}

\DeclareMathOperator*{\argmax}{arg \, max}

\title{Testing isotropy of random fields with values in abstract spaces}

\author[1]{Julius Baumhakel \orcidlink{0009-0006-8999-7303}}
\author[1]{Siegfried Hörmann \orcidlink{0000-0001-5878-2840}}
\author[1]{Matthias Neumann \orcidlink{0000-0002-5782-6773}}

\affil[1]{Institute of Statistics, Graz University of Technology, Austria}

\date{\today}

\begin{document}

\maketitle

\begin{abstract}
We study the problem of testing isotropy, i.e., distributional invariance under orthogonal transformations, for random fields taking values in abstract spaces.  We propose a novel approach based on computing local directional characteristics, which are uniformly distributed under the assumption of isotropy. Our work is motivated by an application in material science, where we are testing the isotropy of function-valued porosity profiles of paper sheets, whose morphology is resolved by µ-CT imaging. While considerably more general, the presented methodology is competitive with existing isotropy tests for real-valued random fields. 
\end{abstract}

\vspace{0.5cm}
\noindent \textbf{Subject Classification:} stat.ME, stat.AP

\section{Introduction}

\subsection{Motivation} 

A random field is stationary if its distribution is shift-invariant, and it is isotropic if its distribution is invariant under any orthogonal transformation (rotation or reflection) at the origin. If the random field is both, stationary and isotropic, we call it motion-invariant. For the analysis of spatial data, the assumption of motion-invariance allows for strong simplifications \citep{cressie_statistics_2015}. In particular, the covariance between two locations depends only on their Euclidean distance. Consequently, several statistical tests have been proposed in the literature to decide whether the assumption of isotropy for random fields is justified by the data. See \citet{weller_review_2016} for an overview of nonparametric tests. In recent years, further tests have been developed, including tests based on the excursion set of a given random field~\citep{abaach_pixel_2025, azais.2025}, spectral domain tests~\citep{weller_nonparametric_2020, eckle_balancing_2026}, and also isotropy tests on the sphere~\citep{sahoo.2019}.

A critical gap is that existing literature addresses primarily real-valued random fields. While multivariate extensions are scarce (see e.g.\ \citet{jona2001modeling}) there is currently a lack of methodology for high-dimensional or non-Euclidean data.  Driven by an application in which we analyze local porosity profiles in paper-based materials represented as point-wise functional data, we aim to bridge this gap by introducing an isotropy test designed for random fields with values in abstract spaces. 
The analysis of local porosity in paper-based materials is of particular interest for industrial applications because it influences effective physical properties such as air permeance~\citep{woudberg.2018, leitl2023joint}. In general, porosity profiles are frequently used descriptors for analyzing the morphology of many functional materials, such as batteries~\citep{hosseinzadeh2018impact, dodell2025quantifying}, fuel cells~\citep{banerjee2016heterogeneous}, or cement~\citep{rimmele2008heterogeneous}.

Also in the context of materials science,  $SO(3)$-valued random fields provide another instance of non-Euclidean random fields, emerging naturally in the 3D imaging of polycrystalline materials~\citep{karafiatova.2022, karafiatova.2023,sheng.2023, rieder2024stochastic}. In this context, the values of the random field represent the crystallographic orientations of the material in space. More generally, tensor- and spinor-valued random fields provide an appropriate mathematical framework to model phenomena at the interface between materials science and continuum physics, as well as in cosmology~\citep{malyarenko.2023}.

\subsection{Dataset}\label{sec:real_data}

The methodology in this paper was developed to test for isotropy of local volume fractions in paper sheets compressed by hard-nip calendering. Local volume fractions are computed based on experimental image data. For this purpose, we consider one 3D image of the data set measured at the PETRA III storage-ring of DESY in Hamburg~\citep{haibel.2010, wilde.2016} described in~\citet{neumann.2021}. This material has been statistically investigated to quantify interdependencies between local morphological descriptors~\cite{Neumann2024} and to elucidate the impact of morphology on air permeance~\citep{Graefensteiner2026}.

\begin{figure}[!htb]
    \centering
    \subfloat[3D rendering of the microstructure.]{
        \includegraphics[width=0.48\linewidth]{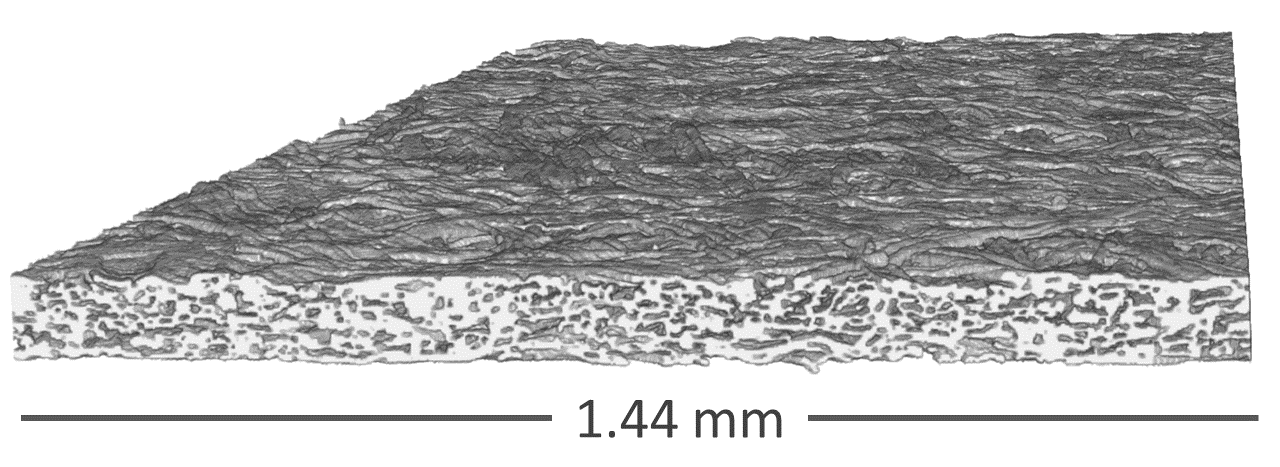}
        \label{fig:real_data_a}
    }
    \hfill
    \subfloat[Definition of functional observations: Vertical slice and example observation at point $x$.]{
        \includegraphics[width=0.48\linewidth]{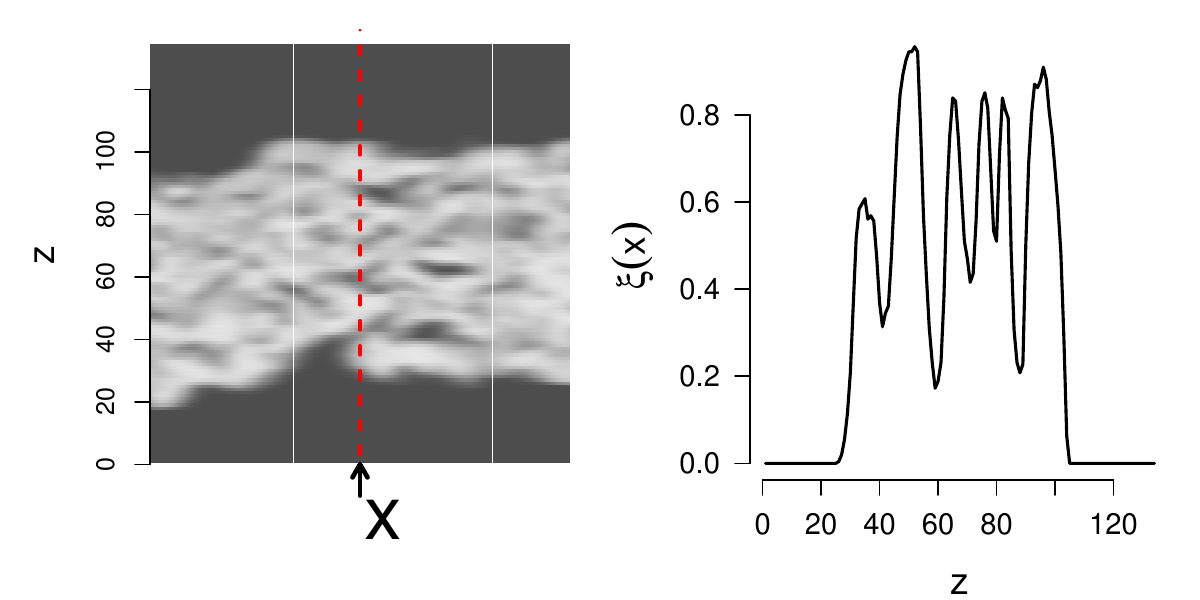}
        \label{fig:real_data_b}
    }
    \caption{3D rendering of the microstructure of a compressed paper sheet, for which we test isotropy of local volume fractions (a), functional observations defined as volume fractions read in thickness direction (b), where the $z$-direction plays the role of $t$ in~\eqref{eq:functionalObservation}.}
    \label{fig:real_data}
\end{figure}

A 3D rendering is provided in Figure~\ref{fig:real_data_a}. The total image consists of $1920 \times 1111 \times 134$ cubic voxels with an edge length of $\mathrm{px} = 1.3~\mu m,$ where each voxel is either labeled as pore or solid. Note that the solid phase is the union of cellulose fibers. The images are smoothed in horizontal direction with a 2D Gaussian kernel with standard deviation $\sigma_{\mathrm G} = 13~\mu m.$ Doing so, we obtain an image, where each voxel contains a number between $0$ and $1$ representing the local volume fraction of cellulose material.

As evident from the 3D rendering in Figure~\ref{fig:real_data_a}, the fibers are concentrated in layers parallel to the surface. When assessing isotropy, we are interested in whether the distribution is invariant under plane-preserving transformations, i.e., rotations or reflections of the paper that do not involve tilting it.
 Hence, we interpret this 3-dimensional image as a realization of a function-valued field on a 2-dimensional domain $\mathcal{D}$,
\begin{equation} \label{eq:functionalObservation}
\xi = \{\xi(x) : x \in \mathcal D\}, \quad   \xi(x)=\{\xi(x)(t)\colon t\in [0,T]\},
\end{equation}
where the $t$ corresponds to the position in through-plane direction observed at some $134$ equidistant points and $\xi(x)$ represents the function of local volume fraction through the paper sheet at position $x$. This interpretation is illustrated in Figure~\ref{fig:real_data_b}.

Testing for isotropy of this function-valued field corresponds to assessing the \textit{planar isotropy} of the material at microscale. For the particular example of function-valued random fields, it is also possible to apply a test developed for spatio-temporal random fields by \citet{li_nonparametric_2007}. However, the latter test assumes stationarity in time, which is not required in the framework developed in the present work and not given in the dataset of interest.

\subsection{Outline of this paper}
The core idea of our work is to recast the problem of testing for isotropy as a problem of testing for uniform distribution in directions. The only structural ingredient we require from the observation space is a pairwise dissimilarity function, formalized by a semi-metric. In normed spaces, this would naturally be the norm of the difference between points.

For each sampling location, we measure the dissimilarity between all points that lie at a fixed distance and record the direction associated with the greatest dissimilarity. We will show that under isotropy, these directions are uniformly distributed. For certain choices of the semi-metric, this procedure has conceptual parallels to methods comparing directional autocovariance, e.g., \citet{guan_nonparametric_2004}.

The remainder of the paper is structured as follows: In Section \ref{s:maintheorems}, we establish the  methodological framework of the proposed hypothesis test. In  Section \ref{s:pract}, we address aspects concerning practical implementation. This includes the estimation of distributional parameters for the test statistic, the choice of hyperparameters, and an outline of the algorithmic procedure. Note that an accompanying R package is also introduced here. Section \ref{sec:finitesample_power} presents an empirical finite-sample power analysis in the scalar-valued case, drawing comparisons to the methods by~\citet{guan_nonparametric_2004} and \citet{weller_nonparametric_2020}. Subsequently, Section \ref{s:application} analyzes the test in the context of a function-valued random field setting, which arises from assessing planar isotropy of the microstructure in paper-based materials. We apply the procedure to a synthetic data set generated from a stochastic fiber model as well as on the real data described in Section~\ref{sec:real_data}.  Finally, Section \ref{sec:discussion} contains concluding remarks and outlines possible future directions. Theoretical justifications of our method are deferred to the Appendix.

\subsection{Notation and setup}  
The object of interest for this paper is a strictly stationary random field $\xi=\{\xi(x)\colon x\in\mathcal{D}\}$ taking values in some measurable space $\mathcal{V}$ and domain $\mathcal{D}\subset b\mathbb{Z}^d$, with sample size $n(\mathcal{D})$. Without loss of generality, we assume that the grid has unit spacing $b = 1$. 

Our methodology can be extended to continuous domains or data observed on irregular spatial locations, but we refrain from developing all three formulations in parallel and leave this open for future works, as our data are sampled on a grid and our algorithms operate in the discrete setting as well. 

The principal target of this work is to develop a statistical test for the hypothesis
$$
H_0\colon \xi\,\, \text{is isotropic}.
$$
Recall that isotropy means that the law of the random field is invariant under rotations and reflections, i.e., if $x_1,\ldots, x_k\in \mathcal{D}$ and if $Q$ is an element of the orthogonal group $O(d)$ such that
$Qx_1,\ldots, Qx_k\in \mathcal{D}$, we have equality of the joint laws, i.e.
\begin{equation}
(\xi(x_1),\ldots, \xi(x_k))\stackrel{\mathcal L}{=}
(\xi(Qx_1),\ldots, \xi(Qx_k)).
\end{equation} 
Along with the required stationarity, the field is said to be \emph{motion invariant}.  

In order to define our test statistic we transform the process $\xi(x)$ into a new field taking values on  $\mathbb H^{(x)}(r)$, a set of points in $\mathcal{D}$ with Euclidean distance $r$ from $x$, i.e. $\mathbb H^{(x)}(r) \subset \mathbb{S}^{d-1}_x(r)\cap \mathcal{D}$ where $\mathbb{S}^{d-1}_x(r)$ denotes the hypersphere in $\mathbb{R}^d$ with radius $r$ and center $x$. 
We suppose that $(\mathcal V, \mathbbl d)$ is an abstract space equipped with a \emph{semi-metric} $\mathbbl d\colon\mathcal{V}\times\mathcal{V}\to [0,\infty)$ which is symmetric in its arguments and satisfies $\mathbbl d(v,w)=0$ if and only if $v=w$ \citep{wilson.1931}. Now we define
\begin{equation}\label{e:m}
M(x,r)\coloneqq \argmax_{y\in \mathbb H^{(x)}(r)} \mathbbl d(\xi(y),\xi(x))\quad \text{and}\quad M_0(x,r)\coloneqq M(x,r)-x,
\end{equation}
which means that $M(x,r)$ identifies the point in $\mathbb H^{(x)}(r)$ at which the field exhibits the largest dissimilarity to $\xi(x)$ within the set of candidates, and $M_0(x,r)$ identifies the lag corresponding to $M(x,r)$.

 If the maximizer in \eqref{e:m} is not unique, we will break ties with a uniform random variable on the set of maximizers, which is independent of everything else. The tiebreakers for different $x$ are also supposed to be mutually independent. 

We fix a basis subset of lags at length $r$ given by $\mathbb H^{(0)}(r) = \{h_1,\ldots,h_m\} \subset \mathbb Z^d \cap \mathbb S_0^{d-1}(r)$, and suppose $\mathbb H^{(x)}(r) = x + \mathbb H^{(0)}(r)$ whenever all required neighbors of $x$ are observed. Further, we denote by $\mathcal D_0$ the set of such $x \in \mathcal D$, where $x + \mathbb H^{(0)}(r) \subset \mathcal D$. $M_0(x,r)$ is thus a field taking values in $\mathbb H^{(0)}(r)$. Examples of the sets $\mathbb H^{(0)}(r)$ as well as $\mathcal D$ and $\mathcal D_0$ are illustrated in Figure~\ref{fig:visualize_sets}.
 
 From now on, we suppose $r$ is fixed, and write $M_0(x)$ and $\mathbb H^{(0)}$ instead of $M_0(x,r)$ and $\mathbb H^{(0)}(r)$.

\section{Methodology}\label{s:maintheorems}

\subsection{Formulation of the test on isotropy}

The key idea behind our method is that under motion invariance, for any fixed point $x$ in the domain, $M_0(x)$ has a uniform distribution on $\mathbb H^{(0)}$ provided the following notion applies to $\mathbb{H}^{(0)}$:

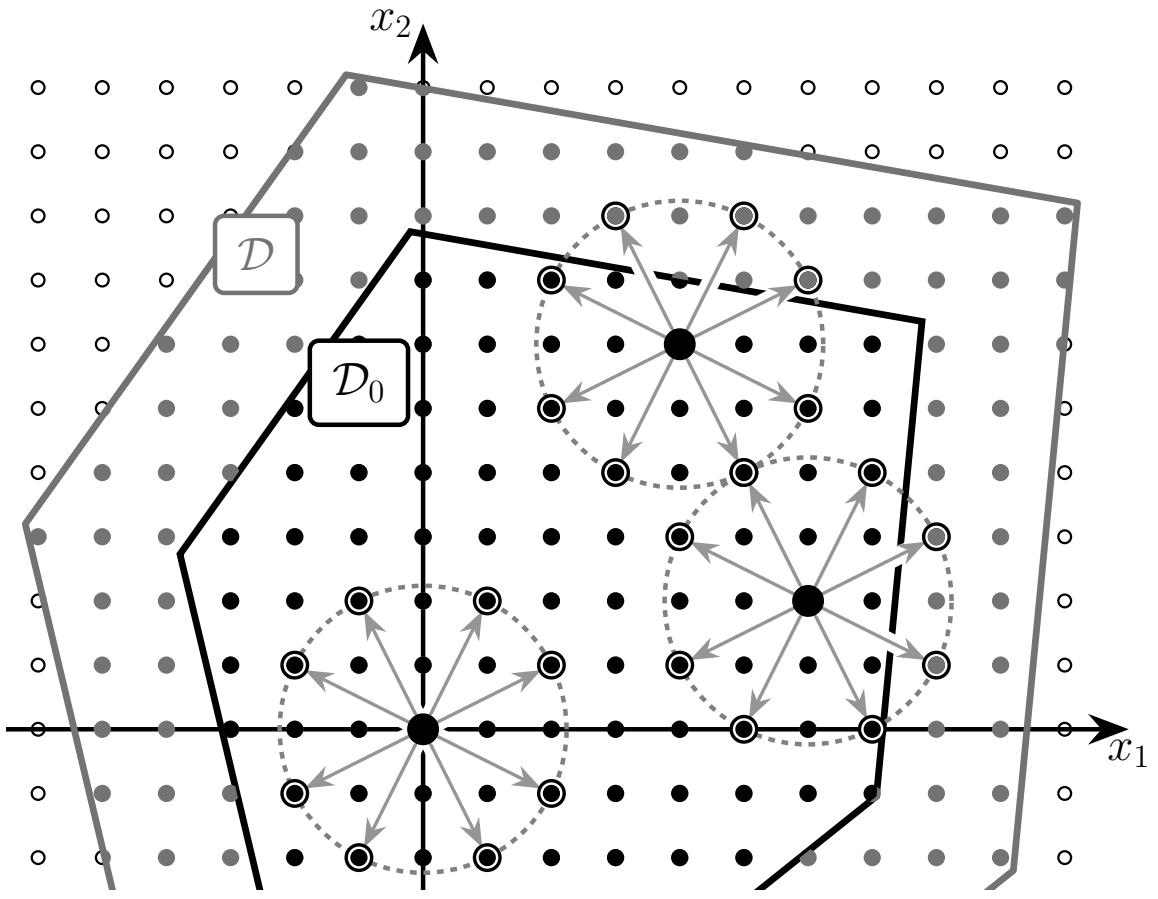
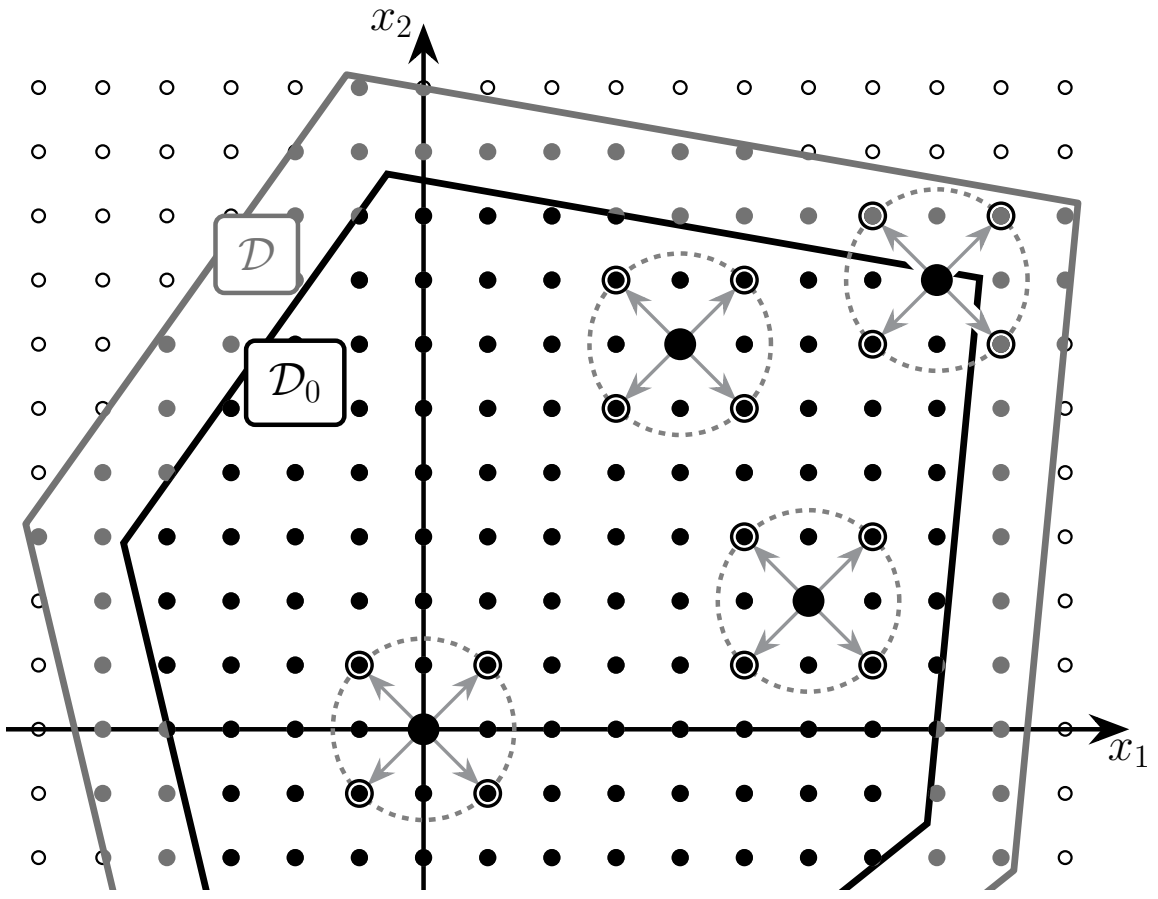
\begin{figure} %
    \centering
    \subfloat[\centering $\mathbb H^{(0)}(\sqrt{5}) = \{ (\pm 2, \pm 1), (\pm 1, \pm 2) \}$]{{\resizebox{0.5\linewidth}{!}{%
    \input{figures/grid_domains}
    }}}
    \subfloat[\centering $\mathbb H^{(0)}(\sqrt{2}) = \{ (\pm 1, \pm 1)\}$]{{\resizebox{0.5\linewidth}{!}{%
    \input{figures/grid_domains_2}
    }}}
    
    \caption{Illustration of the sets $\mathcal D$, $\mathcal D_0$, and $\mathbb H^{(x)}(r)$: The points for which the random field is observed ($\mathcal D$) are drawn in gray and black. The points feasible for further analysis ($\mathcal D_0$)  are shown in black. Moreover, the set of neighbors used in maximizing dissimilarity $\mathbb H^{(x)}(r)$ are illustrated on circles. The arrows indicate lags $\mathbb H^{(0)}(r)$. }
    \label{fig:visualize_sets}
\end{figure}

\begin{definition}\label{def:cyc}
A non-empty set $\mathbb H \subset \mathbb Z^d$ is called \textit{transitive under its orthogonal symmetry group} if there is a generating element $h \in \mathbb H$ so that for every $\tilde h \in \mathbb H$, there exists an orthogonal transformation $Q \in O(d)$ with $Qh = \tilde h$ and $\mathbb H$ invariant under $Q$, i.e. $Q\mathbb H = \mathbb H$.
\end{definition}
This definition is closely related to the algebraic concept of orbits. Specifically, $\mathbb H$ is a transitive orbit of a finite subgroup $G_{\mathbb H}$ of $O(d)$, namely that, which leaves $\mathbb H$ invariant. In the following, we write $A'$ for the transpose of $A \in \mathbb R^{n \times m}$ with $m,n \geq 1.$ 
\begin{example} \label{ex:lags}
    Sets $\mathbb H \subset \mathbb Z^2$ that are transitive under their respective orthogonal symmetry groups include the following:
    \begin{align*}
    &\mathbb{H}_a=\{ (\pm 2, \pm 1), (\pm 1, \pm 2) \},  &&\mathbb{H}_b=\{ (\pm 1, \pm 1)\}, \\
    &\mathbb{H}_c=\{ (\pm 1, 0), (0, \pm 1) \}, && \mathbb{H}_d=\{(1,0), (0,1)\},
    \end{align*}
    where $(\pm a, \pm b)$ denotes the four possible sign configurations. %
    For $\mathbb H \subset \mathbb Z^d$ a general construction with $n(\mathbb H) = 2^d d!$, is the set generated by $h = (1, \ldots, d)'$:
    \[
    \mathbb H = \left\{ (\pm \tau(1), \ldots, \pm \tau(d)) : \tau \text{ is a permutation of } \{1, \ldots, d\}\right\}.
    \]
    Furthermore, all sets $\mathbb H = \{h_1, h_2\}$ with $\norm{h_1} = \norm{h_2}$ are transitive under their respective orthogonal symmetry  (by reflecting along $h_1 + h_2$).
\end{example}

\begin{example} \label{ex:lags2}
    Note that symmetry and equal distance from the origin are not sufficient. E.g., the set
    \[
    \mathbb{H}=\{(\pm 3, \pm4), (\pm4, \pm3), \pm(5, 0), \pm(0, 5)\},
    \]
     is not transitive under its orthogonal symmetry group.%
\end{example}

Some necessary conditions for $\mathbb H^{(0)}$ to be transitive under its orthogonal symmetry group are discussed in Section~\ref{sec:lag_necess_cond} of the Appendix.

The following lemma establishes uniform distribution of $M_0(x)$ under isotropy given the above condition holds for $\mathbb H^{(0)}$. If the condition fails, then $M_0(x)$ will generally not be uniformly distributed.
In Section~\ref{se:counterexample} of the Appendix we give a counter-example.
\begin{lemma}\label{lem:uniformDistributionDiscrete}
    Let $\xi$ be a motion-invariant random field in $\mathbb Z^d$ and let $\mathbb H^{(0)}$ be transitive under its orthogonal symmetry group. Then, for all $x\in \mathcal{D}_0$ we have that $M_0(x)$ is uniformly distributed on $\mathbb H^{(0)}$. 
\end{lemma}
We will now develop a test which assesses the new null-hypothesis $H_0'\supset H_0$ given by
$$
H_0'\colon \text{$M_0(x)$ is uniformly distributed on $\mathbb{H}^{(0)}$}.
$$
A more detailed discussion regarding the relationship between $H_0$ and $H_0'$ is deferred to Section~\ref{se:isounif} below.
We test uniform distribution of $M_0$ by adapting Pearson's $\chi^2$-test for spatially dependent data. To this end, we transform the categorical variables $M_0(x)$ into a dummy vector  $\mathcal I(x)$, whose $i$-th component is equal to 1 if $M_0(x)=h_i$. We denote $\ind_{m} \coloneqq (1,\ldots, 1)'\in\mathbb{R}^m$. Define \begin{equation}\label{eq:latticeCaseTestCLT}
    S \coloneqq  \frac{1}{\sqrt{n({\mathcal{D}_0)}}}\sum_{x \in \mathcal{D}_0} \left[  \mathcal {I}(x) - \frac{1}{m} \ind_{m} \right].
\end{equation}
Let \(\widehat\Gamma\) be some positive semi-definite estimator of the \emph{long-run covariance} matrix given by $\Gamma = \sum_{h \in \mathbb Z^d} \cov \left( \mathcal I(0), \mathcal I(h) \right)$ (see Section~\ref{s:pract}), and suppose \(\widehat F\) denotes the weighted $\chi^2$ distribution with weights given by the eigenvalues of \(\widehat\Gamma\). Define \(T\coloneqq \norm{S}^2\). Our test for isotropy is based on the following decision rule:
\begin{equation}\label{e:decisiond}
    \text{Reject $H_0'$ if}\quad
    T > q_{1-\alpha}(\widehat F),
\end{equation}
where \(q_\alpha(.)\) is the \(\alpha\)-quantile.

Before proceeding, a few remarks are in order:

    \begin{remark}[Asymptotic validity]
     Conditions which guarantee that this is a consistent and asymptotically valid level-$\alpha$ test will be given in the Appendix.
    \end{remark}

    \begin{remark}[Permuting labels]
     If we require   $\widehat\Gamma$ to be affine equivariant, then the test statistic $T$ is also affine equivariant. Particularly, this implies that permuting the labels in $M_0(x)$ (which is replacing $\mathcal{I}(x)$  by $P\mathcal{I}(x)$ with  a suitable  permutation matrix $P$) does not change $T$.  
    \end{remark}

    \begin{remark}[Multiple sets of lags] \label{rem:multipleLagsAllowed}
        While the above methodology only derives a decision rule for rejecting $H_0'$ based on investigating uniform distribution along a single set $\mathbb H^{(0)}(r)$, one may consider multiple such sets $\mathbb H^{(0)}(r_j)$, $j = 1, \ldots, J$. Denote their respective cardinalities by $m_j$ and the corresponding dummy variables by $\mathcal I_j$. Stacking these dummy variables to obtain $\mathcal I^\star \coloneqq (\mathcal I_1', \ldots, \mathcal I_J')'$ and $S^\star$ through centering by $(\frac{1}{m_1} \ind_{m_1}' \ldots \frac{1}{m_J} \ind_{m_J}')'$ gives an analogous decision rule.
    \end{remark}

    \begin{remark}[Testing for reflection symmetry]
        This work focuses on testing for isotropy, i.e., distributional invariance under all orthogonal transformations. However, there are also weaker formulations, like \textit{axial}, \textit{diagonal}, or \textit{complete symmetry}, which limit to invariance under certain reflections only, see \citet{lu_testing_2005} for a hypothesis test of second-order symmetry characteristics. Our approach can be extended to such weaker formulations, given that $\mathbb H^{(0)}$ is transitive under its orthogonal symmetry group limited to the respective reflectional transformations. Similarly, one may test for rotational symmetry only.
    \end{remark}

\subsection{Relationship between $H_0$ and $H_0'$}\label{se:isounif}
We have seen that $H_0\subset H_0'$ and for the latter hypothesis, we have developed a statistical test. In simplified settings it is possible to show that the implication can be reversed, that is  $H_0'\subset H_0$ holds. Evidently, this will depend on the choice of $\mathbbl{d}$ and $\mathbb{H}^{(0)}$. By the generality of our approach,  we cannot expect universally applicable results. Instead, we illustrate the problem on the basis of an example:
\begin{example}\label{ex:ani}
   Let $\xi$ be a stationary univariate Gaussian random field on $\mathbb Z^2$ with geometric anisotropy, meaning that its covariance function has the form
    \begin{equation}\label{eq:ani}
      C(h) \coloneqq \cov(\xi(0), \xi(h)) = \psi(h'Ah),
    \end{equation}
    where $\psi:[0,\infty) \mapsto [0,\infty)$ is strictly decreasing to zero and $A$ is a positive definite matrix.  We set $\mathbb H^{(0)} \coloneqq \{h_1, h_2\}$  with $\norm{h_1} = \norm{h_2}$ as described in Example~\ref{ex:lags}.  
     For $x \in \mathbb Z^2$, let $M_0(x)$ be the lag maximizing $$\mathbbl{d}(\xi(x),\xi(x+h_i)):=\abs{\xi(x + h_i) - \xi(x)},~i \in \{1,2\}.$$ 
    A straightforward computation yields
    \[
    \prob{M_0(x) = h_1} -\frac12=   \frac1\pi \arcsin \operatorname{Corr} ( \underbrace{\xi(h_1) + \xi(h_2) - 2 \xi(0)}_{\eqqcolon U}, \underbrace{\xi(h_1) - \xi(h_2)}_{\eqqcolon V} ).
    \]
    Observing that $\cov(U,V)=2(C(h_2) - C(h_1))$, we conclude that $H_0=H_0'$ unless $C(h_1) = C(h_2)$.
\end{example}
The example shows that in the Gaussian case, geometric anisotropy generally implies a non-uniform distribution of $M_0(x)$ even if $\mathbb{H}^{(0)}$ contains only two lags. In the exceptional case where $C(h_1)=C(h_2)$, we can adapt the test by using a different set  of lags, as evident in the following example. There, we extend the above observation to multivariate fields on higher-dimensional lattices, provided all components share the same geometric anisotropy.
\begin{example}\label{ex:animult}
Let $\xi$ be a stationary Gaussian random field on $\mathbb Z^d$, $d \geq2$, 
with values in $\mathcal V=\mathbb R^p$ so that its matrix-valued covariance function has the form
\begin{equation}\label{eq:animult}
C(h)=\Theta\,\operatorname{diag}\bigl(\psi_1(h'Ah),\ldots,\psi_p(h'Ah)\bigr)\,\Theta',
\end{equation}
where $\Theta\in O(p)$, $A$ is a symmetric positive definite matrix and
$\psi_1,\ldots,\psi_p\colon[0,\infty)\to[0,\infty)$ are strictly decreasing.
Like in the previous example we choose $\mathbb H^{(0)} = \{h_1, h_2\}$ and select
the Euclidean distance for $\mathbbl d$. Then,
\[
\operatorname{sign}\Bigl(\prob{M_0(x)=h_1}-\tfrac12\Bigr)
=\operatorname{sign}\bigl(h_1'Ah_1-h_2'Ah_2\bigr),
\]
Furthermore, there exists a selection of lag sets $\mathbb H^{(0)}_1, \ldots, \mathbb H^{(0)}_{(d-1)(d+2)/2}$, each of cardinality two, so that $M_0(x)$ is uniformly distributed with
respect to all lag sets if and only if $A$ is a positive multiple of the identity matrix, i.e. when $\xi$ is isotropic. 
\end{example}
\begin{remark}
    The construction of the field in Example~\ref{ex:animult} is related to independent component analysis (ICA), where multivariate observations are modeled as linear mixtures of mutually independent latent sources \citet{hyvarinen_independent_2001}. Extensions of this principle to random fields have been considered in the spatial blind source separation literature, see, e.g., \citet{nordhausen_blind_2015}. Our setting corresponds to a special case, where the mixture matrix is orthogonal.
\end{remark}
The claims of the example as well as the construction of the lag sets  $\mathbb H^{(0)}_1, \ldots, \mathbb H^{(0)}_{(d-1)(d+2)/2}$ are further discussed in Section~\ref{sec:appendix_h0_relation} of the Appendix. It supports Remark~\ref{rem:multipleLagsAllowed}: changing lags or adding lags to $\mathbb{H}^{(0)}$ will make the test applicable against a bigger class of alternatives. In the example above, $H_0=H_0'$ if we require uniform distribution of $M_0(x)$ when finding maximal dissimilarity with respect to the lag sets $\mathbb H^{(0)}_1, \ldots, \mathbb H^{(0)}_{(d-1)(d+2)/2}$ simultaneously.

\section{Practical implementation}\label{s:pract}

\subsection{Estimation of the long-run covariance}\label{se:estcov}
Our test statistic is based on the mean of weakly dependent data multivariate, for which \citet{pantle_estimation_2010} introduce a positive semi-definite long-run covariance estimator on a continuous domain. We apply this estimator to the lattice domain case similarly to \citet{sherman_variance_1996} but extended to higher dimensions in a way that preserves mean-square consistency, which we prove in the Appendix.

To compute the estimator, the domain $\mathcal D_0$ is divided into sub-domains, over which the empirical means and their empirical covariance operator are subsequently computed. To this end, select a basis sub-domain $D \subset \mathcal D_0$ (which is, e.g., rectangular) and let $D_{\ell} \coloneqq  D + s_{\ell},~\ell = 1, \ldots, N$ be pairwise disjoint translations of $D$ that fully intersect with the domain $\mathcal D_0$. 

Define
\[
\widehat \mu_{\ell} = \frac{1}{n(D)} \sum_{x \in D_{\ell}} \mathcal I(x) - \frac1m \ind_m\in \mathbb{R}^m, \quad \text{and} \quad \overline \mu \coloneqq  \frac{1}{N} \sum_{\ell = 1}^{N} \widehat \mu_{\ell}.
\]
 The positive semi-definite long-run covariance estimator is then given by
\begin{equation} \label{eq:covEstimator_discrete}
\widehat \Gamma = \frac{n(D)}{N-1} \sum_{\ell=1}^{N} \left( \widehat{\mu}_{\ell} - \overline \mu \right) \left( \widehat{\mu}_{\ell} - \overline \mu \right)^\prime.
\end{equation}
 For extensions to general weight functions and non-disjoint sub-domains, we refer to \citet{pantle_estimation_2010}.

\subsection{User-selected parameters}

\textbf{The semi-metric:} The primary requirement for $\mathbbl d$ is that it must reflect that algebraic or geometric properties of the sample space $\mathcal{V}$ and it should be capable of measuring the specific anisotropy of interest. For example, in the analysis of polycrystalline materials, quotient spaces $SO(3)/G$ of $SO(3)$ play an important role, where $G$ is a subgroup of $SO(3).$ In this case, a physically meaningful semi-metric is \begin{equation*}
    \mathbbl d(v,w) = \min_{g \in G} \arccos \left(\frac{\tr{v^{-1}gw} - 1}{2} \right),
\end{equation*} see Equation~(6) in~\citet{karafiatova.2023}. Here we interpret $v,w$ and $g$ as the matrix representation of the elements in $SO(3).$  For our data example, where $\mathcal{V}=L^2([0,1])$, the Hilbert space of square integrable functions, we will use the standard metric $\mathbbl d(v,w)=\|v-w\|$, where $\|\cdot\|$ is the $L^2$-norm for square integrable functions. %

 It is possible to combine tests for multiple semi-metrics into a single procedure, analogous to the treatment of multiple lags discussed below and in Remark~\ref{rem:multipleLagsAllowed}.\medskip

\noindent
\textbf{The spatial lags:} We need to choose a radius $r$ for the set of lags $\mathbb H^{(0)}(r)$. Obviously, $r\geq 1$, and there must exist integers $j_1,\dots,j_d$ such that $r^2=j_1^2+\cdots+j_d^2$, in order that $\mathbb{H}^{(0)}(r)\neq\emptyset$. Another practical limitation is the size of the domain. 
If the radius is large, the feasible domain $\mathcal{D}_0$ becomes small.

A large radius may also cause vanishing spatial dependence between $\xi(x)$ and $\xi(x+h)$ leading to diminished statistical power. This effect can be seen in Example~\ref{ex:ani}:
if $h_i$ is replaced by $rh_i$ then for $r\to\infty$, $M_0(x)$ becomes uniformly distributed. On the other hand, at small radii, depending on the data situation, the anisotropy may be only barely detectable.

So in practice the choice of $r$ is a tradeoff. In general we propose using multiple smaller lags, and potentially aggregating those into a single hypothesis test (avoiding the need for any a posteriori corrections, such as, e.g., Bonferroni or \v{S}idák corrections), see Remark~\ref{rem:multipleLagsAllowed}.
This procedure is outlined in Algorithm~\ref{alg:combined_test}. 

For the case where $\xi$ is real-valued, as a rule of thumb, we suggest selecting $r$ such that $\operatorname{Corr}(\xi(0),\xi(h)) \approx \frac12$ for $\|h\|=r$, taking $r=1$ if the correlation is already below $\frac12$ at $\|h\|=1$. This can be determined using a direction-independent (isotropic) correlation estimator.

\noindent
\textbf{The basis sub-domain $D$:} The region $D$ must be chosen large enough to adequately cover the long-range dependence of the field. The specific choice of $D$ requires careful consideration, as it directly influences both the size and the power of the test. %
For a detailed derivation of the asymptotically optimal domain size in the two-dimensional setting, see \citet{sherman_variance_1996}.

\subsection{Algorithm}
In the following, we synthesize the concepts from the preceding subsections into a step-by-step computational procedure. Our algorithms are also available in form of an {\tt R} package \citep{baumhakel_baumhakelunisono_2026}.

    \begin{algorithm}
\caption{Hypothesis test for isotropy}
\label{alg:isotropy_test}
\begin{algorithmic}[1]
\State Select distance $r$, lags $\mathbb H^{(0)}(r)$, and semi-metric $\mathbbl d$
\State Determine $\mathcal{D}_0$, the feasible region for optimization
\State For each $x \in \mathcal{D}_0$, find $M_0(x)$ \label{step:loop_start}
\State Compute $S$ \label{step:compute_S} according to Equation~\eqref{eq:latticeCaseTestCLT}
\State Select basis sub-domain $D$ and non-overlapping shifted versions $D_1, \dots, D_N$ \label{step:subsample_start}
\State Compute $\widehat{\Gamma}$ and its eigenvalues $\widehat \lambda_1, \ldots, \widehat \lambda_m$
\State Determine the $(1-\alpha)$-quantile of a weighted $\chi^2$ distribution with weights $\widehat \lambda_1, \ldots, \widehat \lambda_m$ 
\State Compute $T = \norm{S}^2$ and compare it to the quantile\label{step:subsample_end}
\end{algorithmic}
\end{algorithm}

\begin{algorithm}
\caption{Combined hypothesis test for multiple lag sets and semi-metrics}
\label{alg:combined_test}
\begin{algorithmic}[1]
\State Select multiple sets of parameters $(\mathbb H^{(0)}(r_1), \mathbbl d_1), \dots, (\mathbb H^{(0)}(r_J), \mathbbl d_J)$
\State Determine the global feasible region $\mathcal{D}_0$ via the intersection of all individual feasible sets
\For{each parameter set $j = 1, \dots, J$}
    \State Compute $S_j$ by executing Steps~\ref{step:loop_start}--\ref{step:compute_S} of Algorithm~\ref{alg:isotropy_test}
\EndFor
\State Stack $S_1, \dots, S_J$ to form the combined test statistic $S^\star$
\State Execute Steps~\ref{step:subsample_start}--\ref{step:subsample_end} of Algorithm~\ref{alg:isotropy_test} for $S^\star$
\end{algorithmic}
\end{algorithm}

\begin{figure}[!htb]
    \centering
    \subfloat[Compute dissimilarity-maximizing lags (here, candidates are diagonal neighbors).]{
        \includegraphics[width=0.3\linewidth]{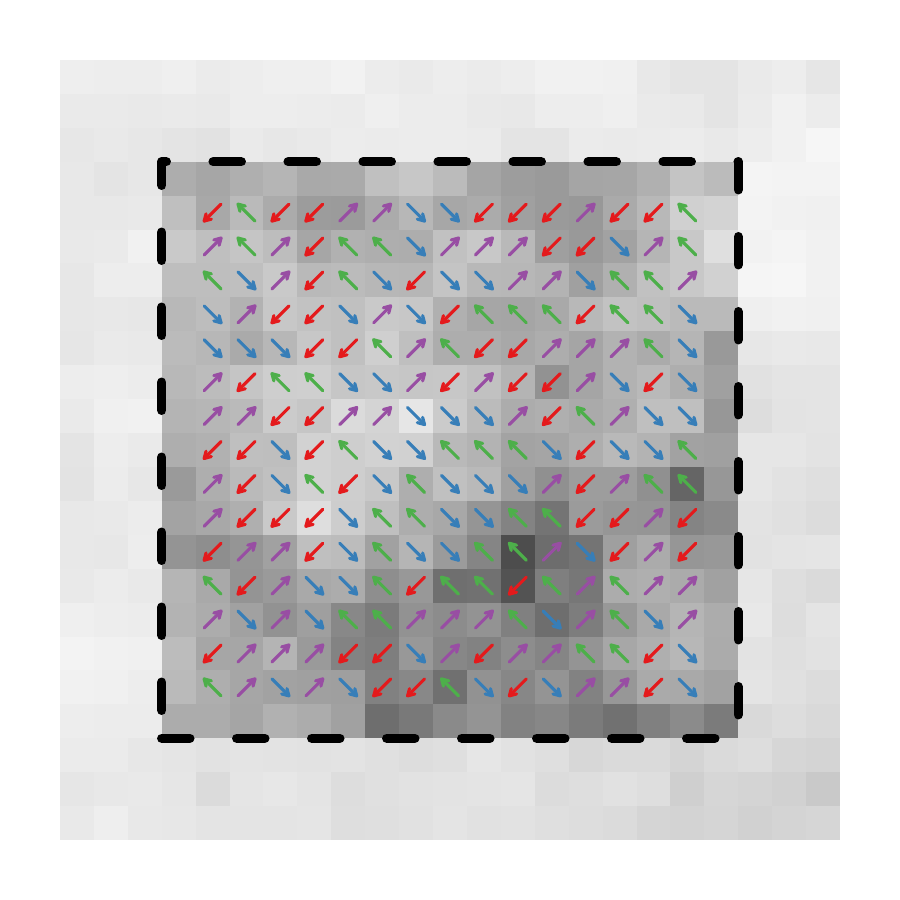}
    }\hfill\makebox[0pt][c]{\tikz[baseline=-0.5ex]{\draw[->, thick] (-0.2,2) -- (0.2,2);}}\hfill
    \subfloat[Compute on whole dataset. Maximizing lags are categorical data.]{
        \includegraphics[width=0.3\linewidth]{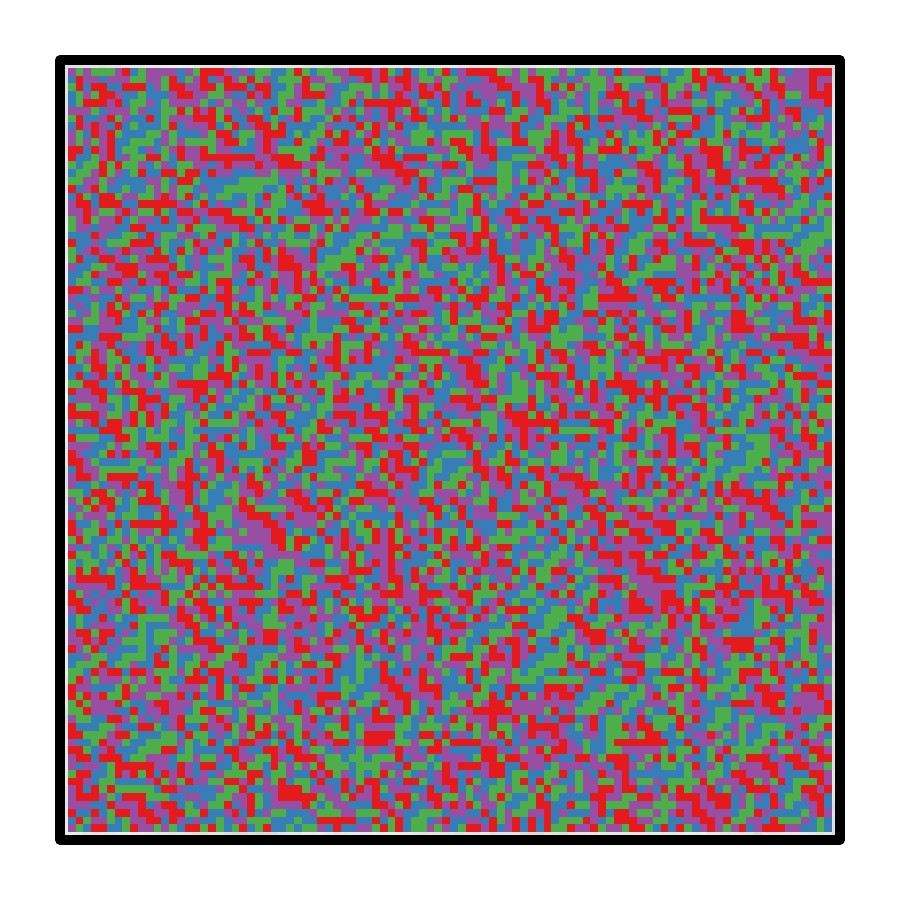}
    }\hfill\makebox[0pt][c]{\tikz[baseline=-0.5ex]{\draw[->, thick] (-0.2,2) -- (0.2,2);}}\hfill
    \subfloat[Compute proportions, compare to uniform distribution.]{
        \includegraphics[width=0.3\linewidth]{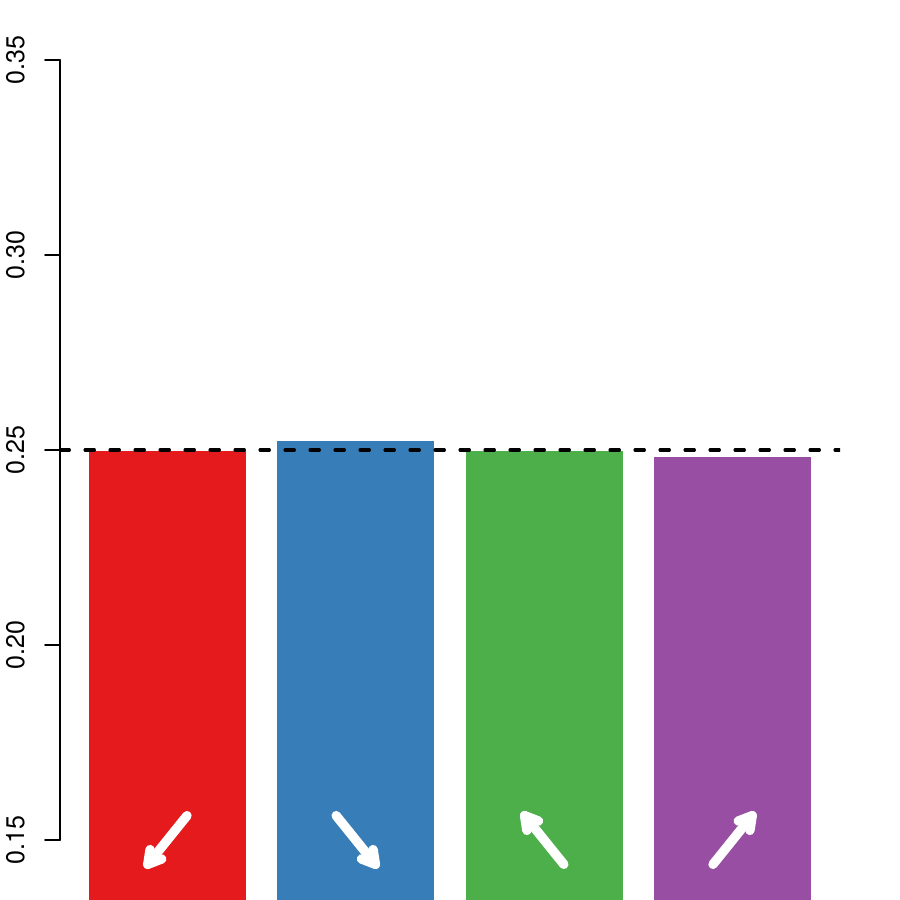}
    }
    \caption{Illustration of Algorithm~\ref{alg:isotropy_test}: Computing dissimilarity-maximizing lags at every grid point (pixel) and comparing proportions to uniform distribution. The example image is a realization of an isotropic Gaussian field.}
    \label{fig:algo_computation}
\end{figure}

Algorithm~\ref{alg:isotropy_test} outlines the testing procedure for a single set of lags and a single semi-metric. The computations in this basic algorithm are visually summarized in Figure~\ref{fig:algo_computation}, which illustrates the local computation of maximizing lags at the grid level, and global aggregation. Furthermore, Algorithm~\ref{alg:combined_test} details how this approach naturally extends to a combined test for multiple sets of lags and semi-metrics by appropriately stacking the test statistics. To determine the quantiles of the weighted $\chi^2$ distribution in the final step of both algorithms, we suggest Davies' method \citep{davies_algorithm_1980}.

\section{Finite sample behavior and power analysis}

\label{sec:finitesample_power}
While the primary motivation of our work lies in accommodating high-dimensional and functional data objects, we will now evaluate its finite-sample performance in the scalar setting and compare to specialized tests. Several nonparametric tests for spatial isotropy have been proposed for scalar-valued fields (see e.g. \citet{weller_review_2016}), and we benchmark the empirical size and power of our approach against some of these established baselines.

To this end, we replicate the simulation experiment detailed by \citet{weller_nonparametric_2020}. Specifically, we compare our proposed test (\textbf{UNI}, Uniform Neighborhood Isotropy testing) to the spectral domain test developed by \citet{weller_nonparametric_2020} (\textbf{SIS}%
) and the covariance-based test by \citet{guan_nonparametric_2004} (\textbf{GSC}). Note that the GSC test also requires to select a set of spatial lags $\mathbb{H}^{(0)}$ with equal lengths on which it tests for equality of spatial covariance as well as a subsampling window, while SIS does neither.

As data generating process we consider a centered, stationary Gaussian random field $\xi=\{\xi(x) : x \in \mathcal{D} \subset \mathbb{Z}^2\}$ and draw realizations on a regular grid. The spatial dependence is governed by a Matérn covariance function, given by
\begin{equation}
    C(h) = \frac{\sigma^2}{2^{\kappa - 1} \Gamma(\kappa)} \left( \frac{\norm{A(\theta, R) h}}{\phi} \right)^\kappa K_\kappa \left( \frac{\norm{ A(\theta, R) h }}{\phi} \right),
\end{equation}
where $\sigma^2 > 0$ is the marginal variance, $\phi > 0$ is the range parameter, $\kappa > 0$ is a parameter related to smoothness, and $K_\kappa$ is the modified Bessel function of the second kind of order $\kappa$. The linear transformation matrix $A(\theta, R)$ induces geometric anisotropy and is given by
\begin{equation}
    A(\theta, R) = \begin{pmatrix} 1 & 0 \\ 0 & 1/R \end{pmatrix} \begin{pmatrix} \cos\theta & -\sin\theta \\ \sin\theta & \cos\theta \end{pmatrix},
\end{equation}
where $\theta \in [0, \pi/2)$ is the angle of the principal axis of anisotropy and $R \geq 1$ is the anisotropy ratio. Starting out with an isotropic field, those parameters govern in which direction ($\theta$) and how much ($R$) the field is stretched in order to obtain the anisotropic one. 

Following the specifications of \citet{weller_nonparametric_2020}, we set the covariance parameters to $\sigma^2 = 1$, $1/\phi = 1.52$, and $\kappa = 1$. The fields are generated and evaluated on different square grids to assess the effect of the sample size: $15 \times 15$ up to $50 \times 50$ in $5$-step increments for each side length.

The null hypothesis of strict spatial isotropy corresponds to $R = 1$, under which the angle $\theta$ is unidentifiable (set to $\theta = 0$ for convenience). To evaluate the empirical power, we simulate fields under combinations of anisotropy ratios $R \in \{\sqrt{2}, 2\}$ and rotation angles $\theta \in \{\pi/8, \pi/4\}$.

We set GSC up to test for equality of autocovariance along the lags $(1,0)$ and $(0,1)$, as well as $(1,1)$ and $(-1,1)$. We further choose subwindow dimensions for long-run covariance estimation as $2 \times 2$. Due to constraints in computation time, GSC is not computed for domains larger than $30 \times 30$. Regarding UNI, we similarly test for uniform distribution along lags $\{(\pm1, 0), (0, \pm1)\}$, and $\{(\pm1, \pm1)\}$. For estimating the long-run covariance matrix, we set the subwindow size to $4 \times 4$ here. While SIS has no user-defined parameters, square grids lead to inflated size, which is why instead of computing the statistic on the full $\ell \times \ell$ grid, we compute it on $\ell \times (\ell-1)$ data points only.

\begin{figure}[!htb]
    \centering
    \subfloat[Proposed test (UNI).]{
        \includegraphics[width=0.48\linewidth]{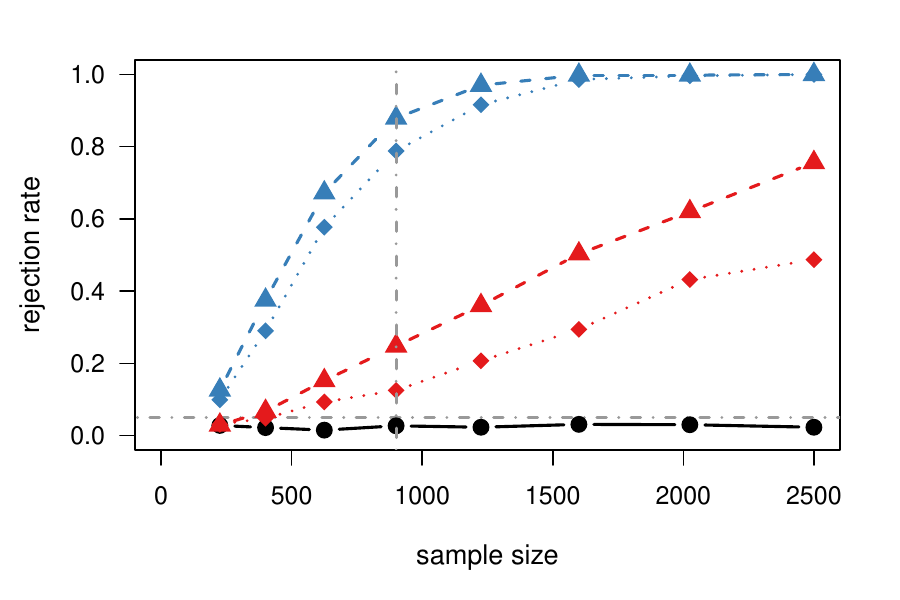}
        \label{fig:gausspower_uni}
    }
    \hfill
    \subfloat[\citet{weller_nonparametric_2020} (SIS).]{
        \includegraphics[width=0.48\linewidth]{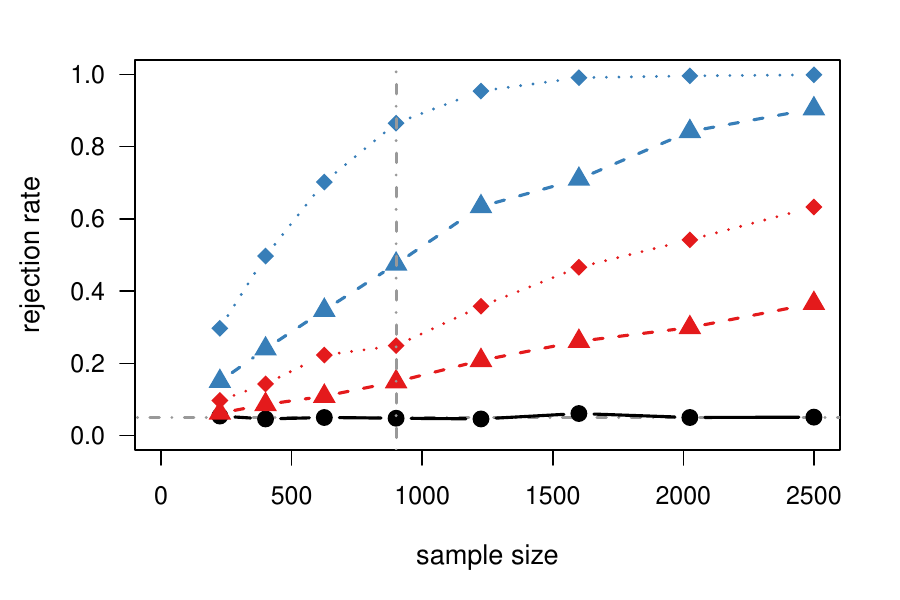}
        \label{fig:gausspower_sis}
    }
    
    \subfloat[\citet{guan_nonparametric_2004} (GSC).]{
        \includegraphics[width=0.48\linewidth]{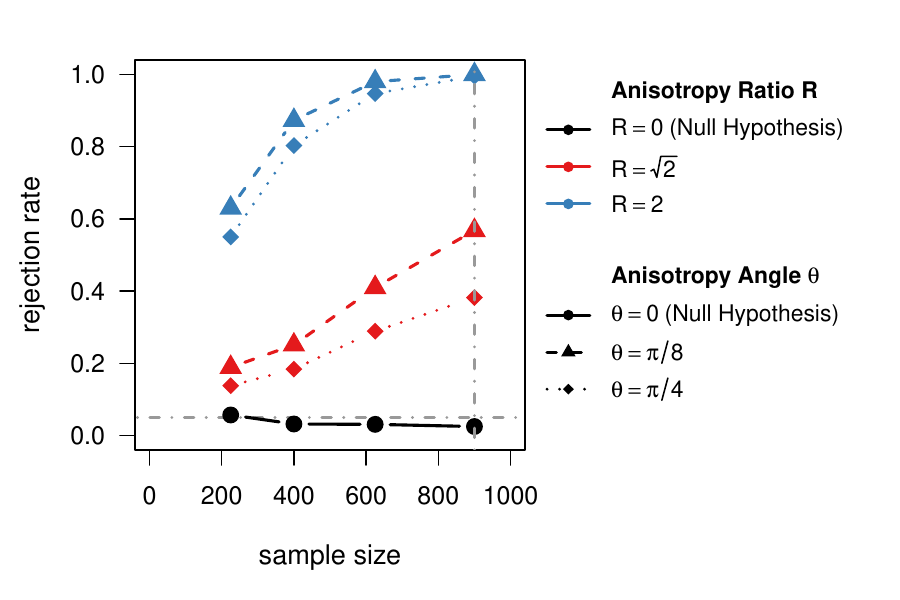}
        \label{fig:gausspower_gsc}
    }
    \hfill
    \subfloat[Average computation time in $\log_{10}(\text{seconds})$.]{
        \includegraphics[width=0.48\linewidth]{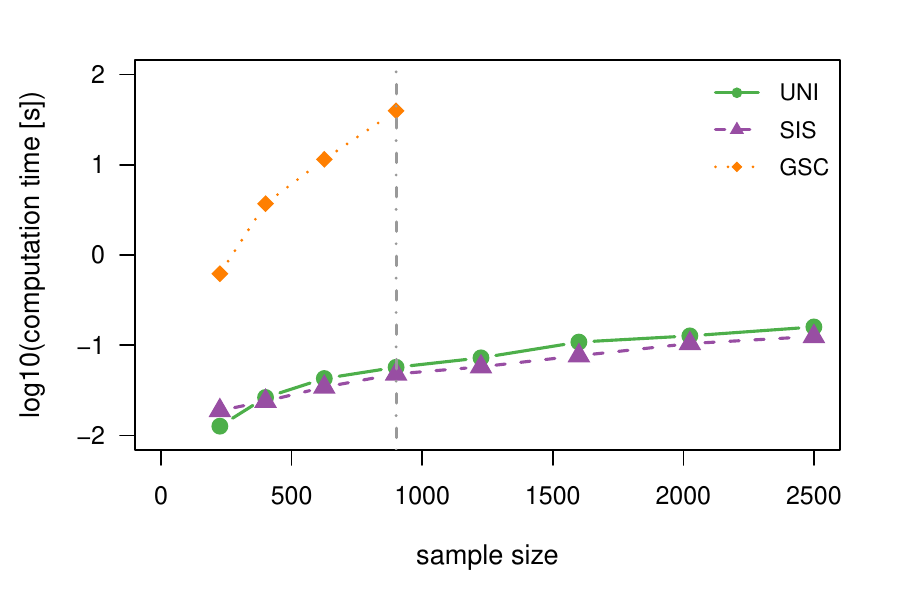}
        \label{fig:gausspower_comptime}
    }
    \caption{Empirical size and power as well as average computation time for testing spatial isotropy at the $\alpha = 0.05$ significance level. Results are based on $1000$ Monte Carlo simulations for various anisotropy ratios ($R$) and angles ($\theta$). Sample size is increased in steps of $5$ units per direction, from a grid of size $15 \times 15$ up to $50 \times 50$. GSC is only computed up to side length $30$ due to computation time. The corresponding sample size up to which GSC is computed is marked with a vertical gray line in each image.}
    \label{fig:gausspower_sizeincrease}
\end{figure}

All tests are run in R \citep{r_core_team_r_2023}. For GSC, the implementation of the package \texttt{spTest} \citep{weller_sptest_2018} is used. The empirical size and power results at the nominal $0.05$ significance level as well as computation times for selected sample sizes are summarized in %
Figure~\ref{fig:gausspower_sizeincrease}.

In the investigated scalar setting, UNI and SIS exhibit similar statistical power and computation times. GSC, in contrast, achieves higher statistical power, but is substantially more expensive in computation: depending on sample size, its computation time is up to two orders of magnitude larger than that of UNI and SIS. This additional computational cost is mainly due to the more involved estimation of the long-run covariance. Moreover, both GSC and UNI require the selection of a specific set of lags and a subwindow size. SIS instead considers all lags simultaneously and avoids long-run covariance estimation via subsampling. However, this computational simplicity comes at a statistical cost, as SIS exhibits inflated test size for square grids. Overall, these results indicate that the proposed UNI test is competitive with existing methods even in the scalar setting, despite it being applicable much greater generality.

\section{Application to fiber-based materials} \label{s:application}

Before applying our UNI procedure to the dataset, we investigate its size and power empirically on synthetic fiber data, via a stochastic fiber model.  With the fiber model, we mimic porous structures of a similar type as the data.

\subsection{Stochastic fiber model} \label{se:stoch_fiber_model}

We use a slightly modified version of the soft-core fiber model introduced in Section II.A of~\citet{altendorf.2011}. The solid phase is modeled by a Boolean model $\Xi$, see Chapter~6 in~\citet{jeulin.2021}, with fibers as grains. For this purpose, let $P_1, P_2, \ldots$ be a homogeneous Poisson point process in $\mathbb R^2 \times [0,K]$ with intensity $\eta > 0$~\citep{lastpenrose2018}. Moreover, let $\Xi_1, \Xi_2, \ldots$ be a sequence of i.i.d.\ copies of the random compact set $\Xi_0$, defined as follows. For a fixed $n_{\mathrm b} \in \mathbb N,$ we define a random sequence $\{p_0, p_1, \ldots, p_{n_{\mathrm b}}\},$ where $p_0 = 0$ is the origin in $\mathbb R^3$. Then \begin{equation}\label{eq:formula_individual_fibers}
    \Xi_0 = \bigcup_{i=0}^{n_{\mathrm b}} p_i + B_{r_{\mathrm f}}(0),
\end{equation}
where $B_{r_{\mathrm f}}(0)$ denotes the closed ball with radius $r_{\mathrm f} > 0$ centered at the origin. Here, $r_{\mathrm f} > 0$ is a model parameter that describes the thickness of the fibers. The random sequence $p_1, \ldots, p_{n_{\mathrm b}}$ models the medial axis of a fiber with radius $r_{\mathrm f}>0.$ In the following, we describe the distribution of this sequence. First, the main orientation $\mu_0 \in \mathbb{S}^2$ is modeled by a Bingham distribution, see Section 9.4.3 in~\citet{mardia_directional_2000}, i.e. the probability density function $f_{\mu_0}:\mathbb S^2 \rightarrow[0,\infty)$ of  $\mu_0$ is given by 
\begin{equation} \label{eq:binghamdens}
    f_{\mu_0}(x) = c_B \, \exp(x' B x),
\end{equation}
for each $x \in \mathbb S^2$, where $B \in \mathbb R^{3\times3}$, positive semi-definite, is a model parameter and $c_B > 0$ denotes the normalizing constant. Then, we model $p_1 = r_{\mathrm f} \mu_0/2.$ The remaining $p_2, \ldots, p_{n_{\mathrm b}}$ are modeled by $p_j = p_{j-1} + \mu_{j-1}  r_{\mathrm f}/2$, where conditional on $\mu_0$, the directions $\mu_1, \ldots, \mu_{n_{\mathrm b}-1}$ form a Markov chain in $\mathbb S^2$. For each $2 \leq j \leq n_{\mathrm b}$, the conditional probability density function is
\begin{equation*}
    f_{\mu_j \mid \mu_{j-1}}(x) = \frac{|\kappa_1 \mu_0 + \kappa_2 \mu_{j-1}|}{2\pi (\exp(|\kappa_1 \mu_0 + \kappa_2 \mu_{j-1}|) - \exp(-|\kappa_1 \mu_0 + \kappa_2 \mu_{j-1}|))} \exp(\kappa_1 \mu_0' x + \kappa_2 \mu_{j-1}'x)
\end{equation*}
for $x \in \mathbb S^2$. The parameters $\kappa_1, \kappa_2 > 0$ are  controlling the influence of the main direction $\mu_0$ and the previous direction $\mu_{j-1}.$ Note that $\mu_j$ follows a von Mises-Fisher distribution conditioned on $\mu_{j-1},$ see Section 9.3.2 in~\citet{mardia_directional_2000}. Finally, the entire sequence $\{0, p_1, \ldots, p_{n_{\mathrm b}}\}$ is rotated around the origin, such that the main direction of the fiber, which is $p_{n_{\mathrm b}} / |p_{n_{\mathrm b}}|,$ coincides with $\mu_0.$ The latter is described in detail in Section II.A of~\citet{altendorf.2011}. With this definition of the medial axis, we have specified the model of individual fibers given in~\eqref{eq:formula_individual_fibers}. The solid phase, which is the union of individual fibers, is then defined by
\begin{equation}\label{eq:boolean_model_fibers}
    \Xi = \bigcup_{i=1}^{\infty} P_i + \Xi_i.
\end{equation}

For the Boolean model in~\eqref{eq:boolean_model_fibers}, the expected volume fraction is known, i.e. 
\begin{equation}
    \frac1{\lambda(W)}\mathbb E [\lambda(\Xi \cap W)] = 1 - \exp(-\eta \, \mathbb E [\lambda(\Xi_0)]), \label{eq:volume_fraction_Boolean_model}
\end{equation}

for each Borel set $W \subset \mathbb R^3,$ see Equation~(3.15) in~\citet{chiu.2013}. The resulting set $\Xi$ is stationary and $m$-dependent where $m = 2r_{\mathrm f} (n_{\mathrm b} + 2),$ see Equation~(3.18) in \citet{chiu.2013} and thus fulfills weak dependence constraints required for asymptotic validity of the test, discussed in the Appendix.

For the power study, we draw realizations of the fiber model that are similar to the data set introduced in Section~\ref{sec:real_data} and represent the microstructure of a compressed paper material. Due to the compression, the fiber cross sections have elliptical shape instead, which is why we consider the transformed random closed set 
$$\Xi_\zeta = \{(x_1, x_2, x_3) \in \mathbb R^3: (x_1, x_2, \zeta x_3) \in \Xi\}$$
for a compression factor $\zeta = 0.4$. Note that $\Xi_\zeta$ is again stationary and has the same volume fraction as $\Xi$.

The simulations are performed on a voxel grid, where the voxels are cubic with an edge length of $\mathrm{px} = 6.5~\mu m$. For simplicity, we simulate from $\Xi$ instead of $\Xi_{0.4}$ with fiber length $L = 140~\mathrm{px}$ and cross-sectional radius $r_{\mathrm f} = 2.5~\mathrm{px}$. Finally, we adapt the edge lengths of the voxels accordingly to obtain $\Xi_{0.4}$. Moreover, note that our simulations model the interior part of fiber-based materials, while the image data of paper-based materials also show the boundaries of the paper, see Figure~\ref{fig:real_data_a}. We further reduce the resolution of the discretized images by a factor of 5. Altogether, this means that $\Xi_\zeta$ is evaluated on a $400 \times 200 \times 30$ grid.

To determine the number of distinct balls $n_{\mathrm b}$ making up the fiber, we use
\[
L=r_{\mathrm f} \frac{n_{\mathrm b}+4}{2}.
\]
In order to imitate the volume fraction observed in image data of approximately $\rho_0 = 160/255$, we estimate the intensity $\eta$ by means of Equation~\eqref{eq:volume_fraction_Boolean_model}, where the volume fraction of the left-hand side is estimated from image data and the expected volume fraction of an individual fiber is computed by means of Monte Carlo simulations.

 By the spectral theorem we can decompose $B = Q\Lambda Q'$ in \eqref{eq:binghamdens}, where $Q$ is an orthogonal, and $\Lambda$ a diagonal matrix. 

We parametrize $Q = Q_\theta$ and define it to be the matrix representation of rotation around the $y$-axis by angle $\theta$, and $\Lambda = \Lambda(c) = \operatorname{diag}(100, 100-c, 0)$. For $c = \theta = 0$, the resulting Bingham distribution concentrates around the equator of the sphere. Increasing $c$, the mass concentrates on two antipodal points on the equator. When $\theta$ is chosen away from $0$, this circular area of concentration tilts along the $y$-axis. 

\begin{figure}[!htbp]
    \centering
    \newcommand{\colOneWidth}{0.3\textwidth}
    \newcommand{\colTwoWidth}{0.7\textwidth}
    
    \subfloat[Bingham density \\ $(c = 0$, $\theta = 0)$.\label{fig:bingham_0_0}]{%
        \begin{minipage}[b]{\colOneWidth}
            \centering
            \includegraphics[width=\textwidth]{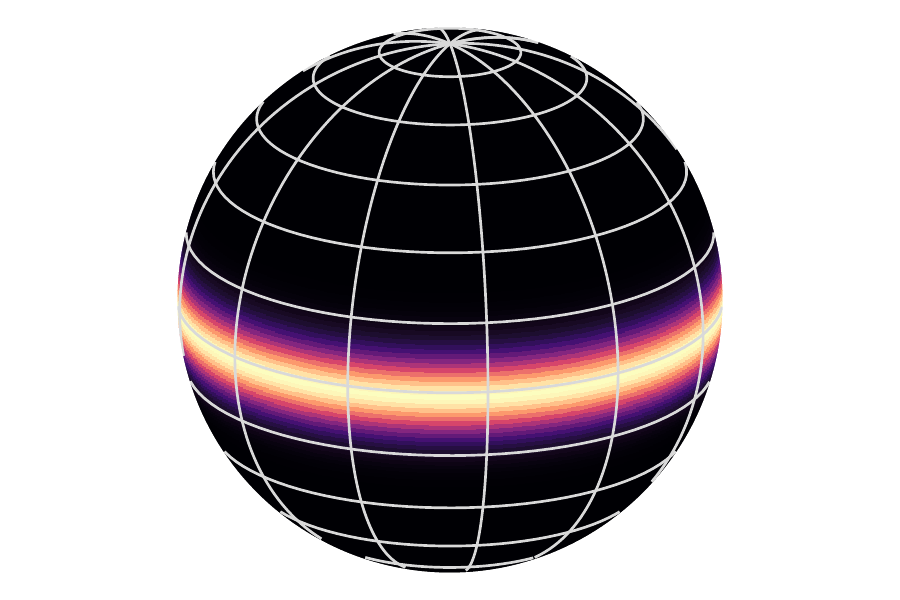}
        \end{minipage}%
    }%
    \hfill %
    \subfloat[Slices of 3-dimensional structures simulated with the stochastic fiber model $(c = 0$, $\theta = 0)$. Dashed lines indicate observation windows $100 \times 50$ (red), $200 \times 100$ (blue), $400 \times 200$ (green). Fibers run mostly horizontally, field fulfills planar isotropy.  \label{fig:fibers_0_0}]{%
        \begin{minipage}[b]{\colTwoWidth}
            \centering
            \includegraphics[width=\textwidth]{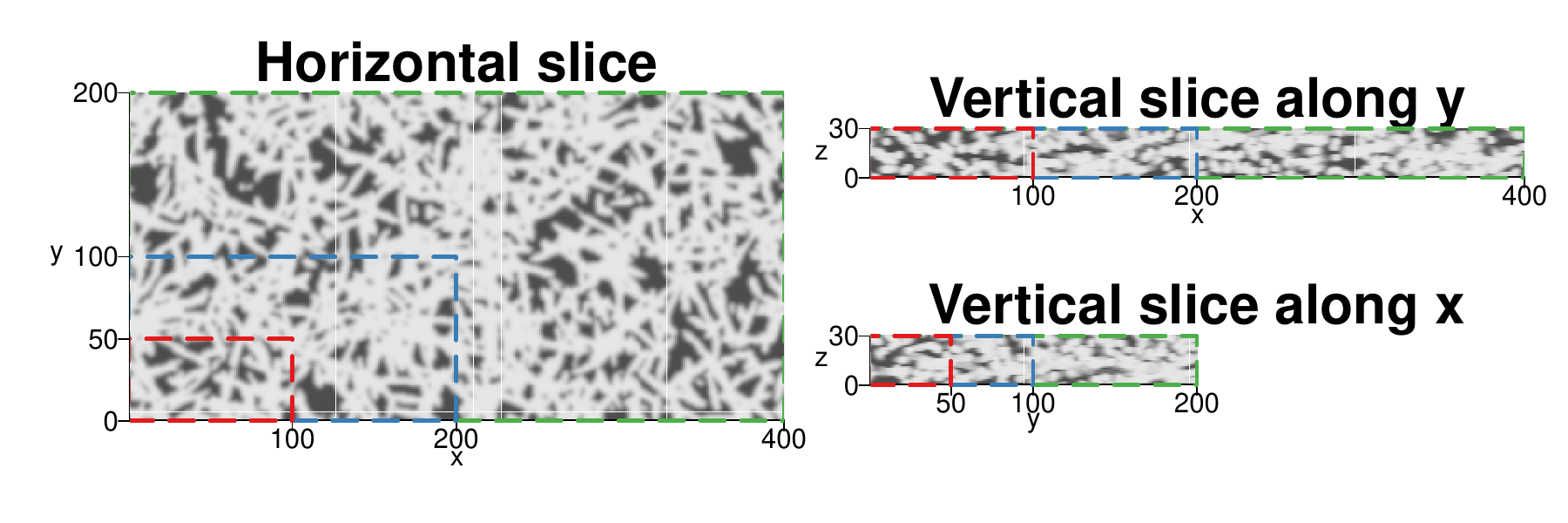}
        \end{minipage}%
    }
    
    \vspace{\floatsep} %
    
    \subfloat[Bingham density \\ $(c = 5$, $\theta = 0)$.\label{fig:bingham_5_0}]{%
        \begin{minipage}[b]{\colOneWidth}
            \centering
            \includegraphics[width=\textwidth]{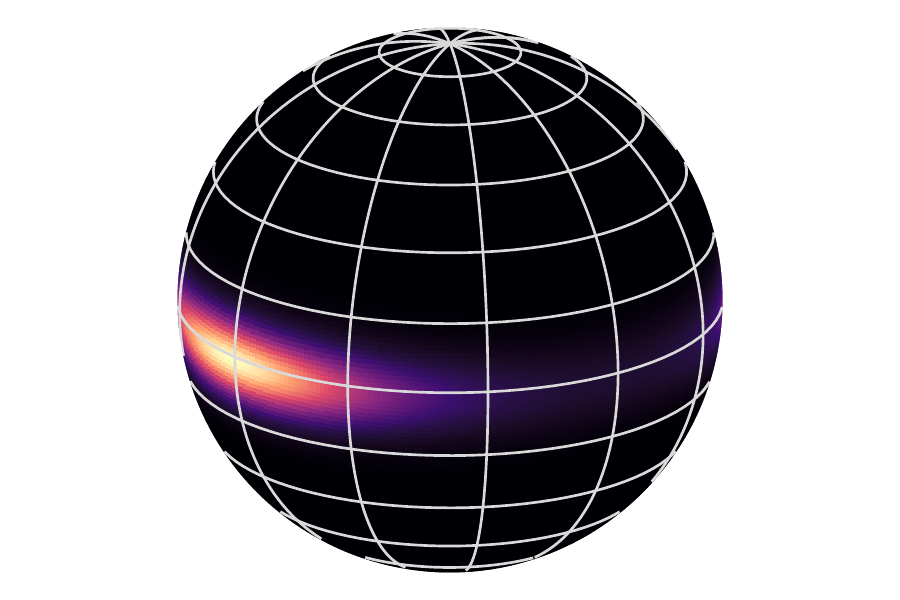}
        \end{minipage}%
    }%
    \hfill
    \subfloat[Slices of 3-dimensional structures simulated with the stochastic fiber model $(c = 5$, $\theta = 0)$. Fiber directions are concentrated to preferably extend in $x$-direction.\label{fig:fibers_5_0}]{%
        \begin{minipage}[b]{\colTwoWidth}
            \centering
            \includegraphics[width=\textwidth]{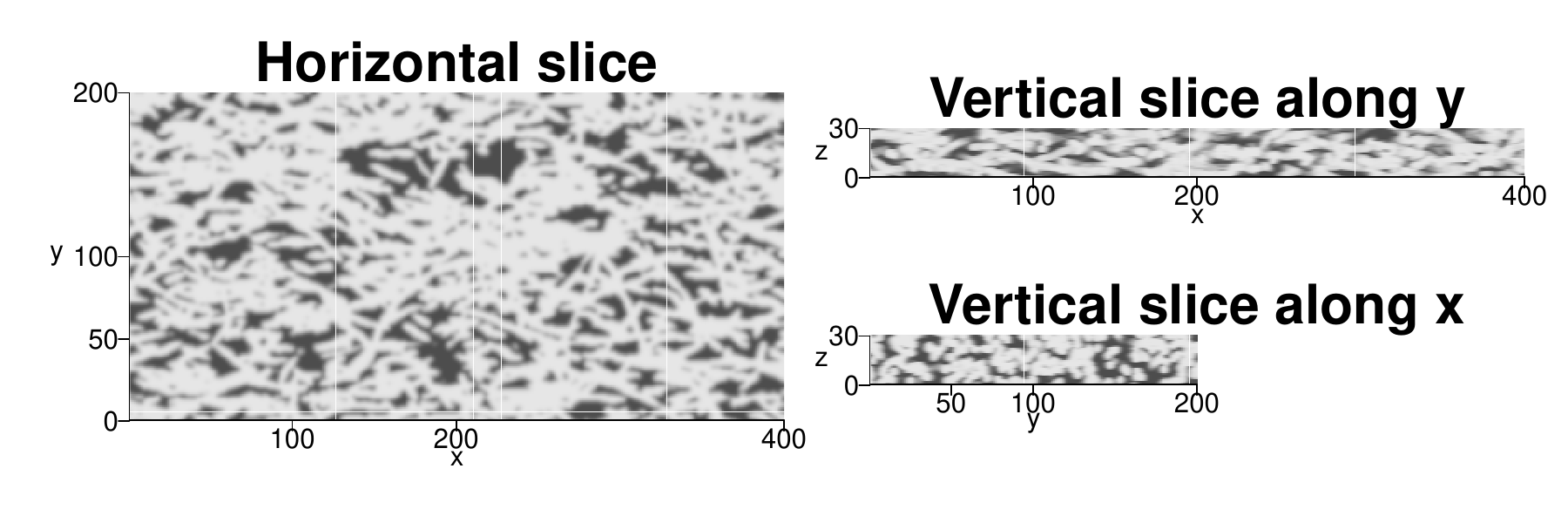}
        \end{minipage}%
    }
    
    \vspace{\floatsep} 
    
    \subfloat[Bingham density \\ $(c = 0$, $\theta = \pi/6)$.\label{fig:bingham_0_30}]{%
        \begin{minipage}[b]{\colOneWidth}
            \centering
            \includegraphics[width=\textwidth]{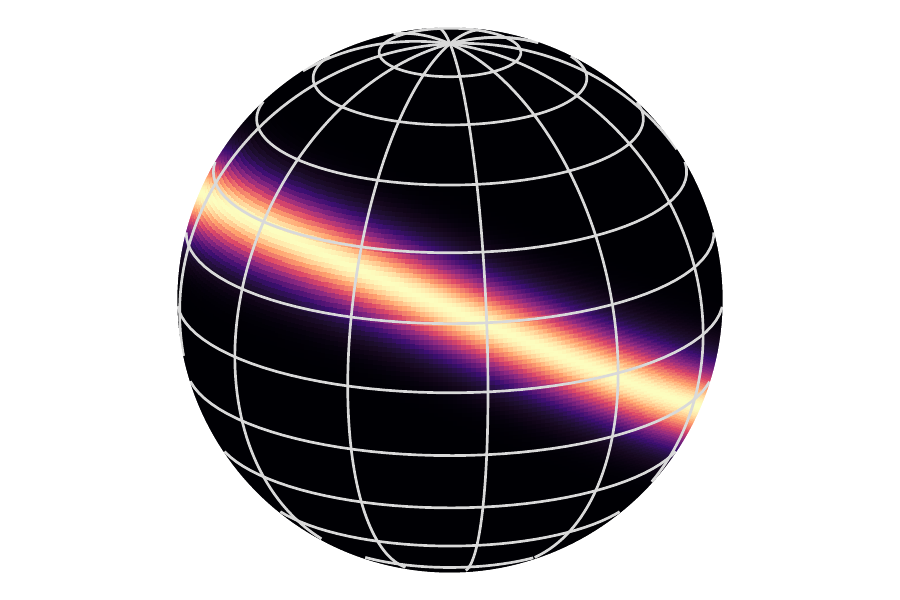}
        \end{minipage}%
    }%
    \hfill
    \subfloat[Slices of 3-dimensional structures simulated with the stochastic fiber model $(c = 0$, $\theta = \pi/6)$. Fiber directions are mostly horizontal if they run in $y$-direction, but have slope $\theta = \pi/6$ if they run in $x$-direction.  \label{fig:fibers_0_30}]{%
        \begin{minipage}[b]{\colTwoWidth}
            \centering
            \includegraphics[width=\textwidth]{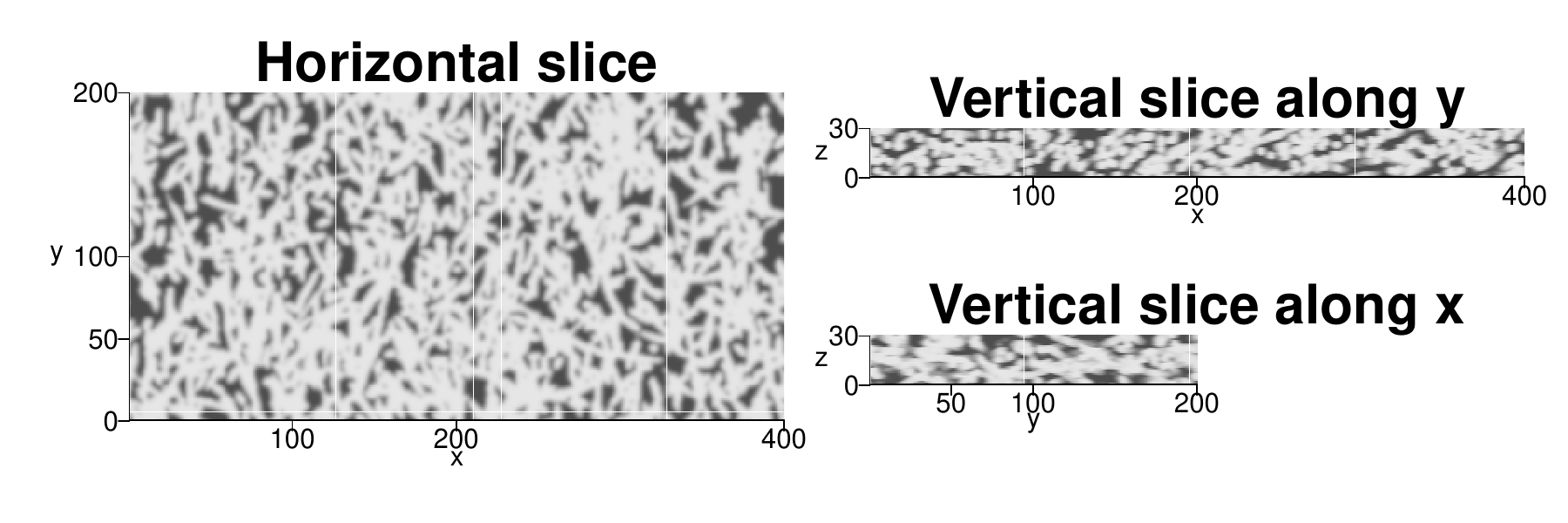}
        \end{minipage}%
    }

    \caption{Illustration of structures simulated with the stochastic fiber model in case of the null hypothesis (first row), and the extreme cases of the respective alternatives (second and third row). The first column illustrates the density of the Bingham distribution used to sample the main direction of each fiber. The second column shows one instance of the generated fibers, respectively, given by slicing the 3-dimensional image across the three cardinal directions. Units are given in pixels.
    \label{fig:fiber_slices}}
\end{figure}

We thus obtain a setting of planar isotropy for $c = \theta = 0$ and two ways of inducing anisotropy, where in the first situation, we increase $c$ from $0$ up to a maximum of $c = 5$, and for the second, $\theta$ from $0$ up to $\pi/6$. The resulting densities for these cases as well as exemplary realizations of the stochastic fiber model are shown in Figure~\ref{fig:fiber_slices}.

We investigate the power and size empirically on three sizes of observational windows: $100 \times 50$, $200 \times 100$, and $400 \times 200$, the latter corresponding to the downscaled dimensions of the dataset. The extent in $z$-direction is fixed to $30~\mathrm{px}$. Figure~\ref{fig:fibers_0_0} illustrates these window sizes relatively to the generated fiber images.

\subsection{Results}

\begin{figure}[!htb]
    \centering
    \subfloat[Sets of lags used to evaluate isotropy of stochastically generated images. Lengths are upscaled for real dataset.
    \label{fig:fiber_lags}]{%
        \begin{minipage}[b]{0.38\textwidth}
            \centering
            \includegraphics[width=\textwidth]{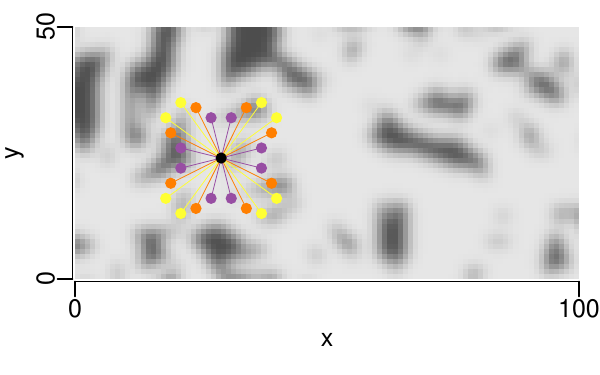}
        \end{minipage}%
    }
    \hfill
    \subfloat[Proportion of selections and joint $95\%$ asymptotic confidence intervals for each lag for paper data. Angle of radial lines corresponds to angle of lag. Length of radial lines is given by selection proportions. Red circle corresponds to the uniform distribution ($1/8$). Ring sections outline confidence intervals for the proportions (in radial direction). Generated by R package unisono \citep{baumhakel_baumhakelunisono_2026}.\label{fig:fiber_lag_selection_proportions}]{%
        \begin{minipage}[b]{0.58\textwidth}
            \centering
            \includegraphics[width=\textwidth]{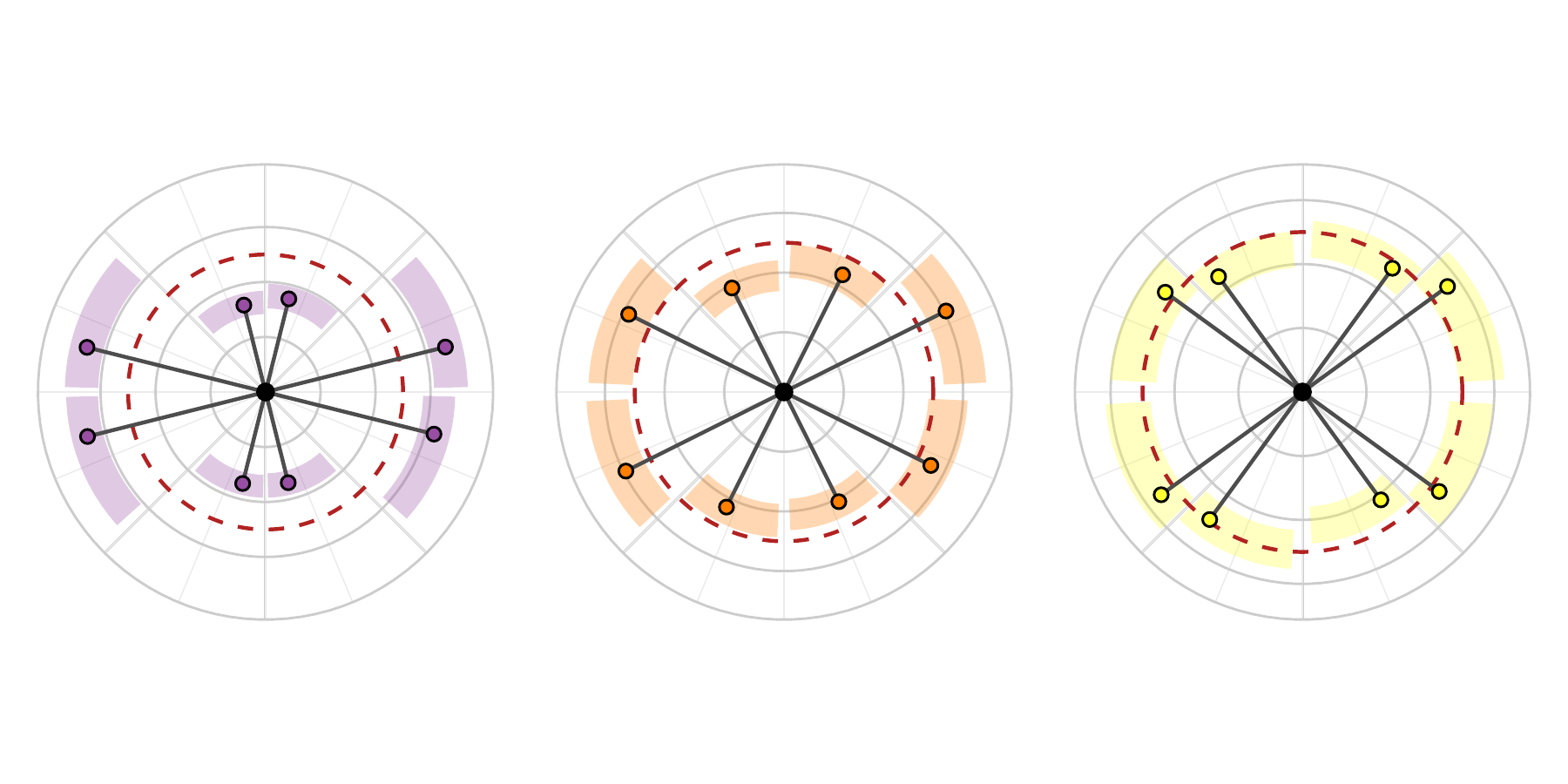}
        \end{minipage}%
    }
    \caption{Illustration of sets of lags $\mathbb H_1$ (purple), $\mathbb H_2$ (orange), and $\mathbb H_3$ (yellow) and selection proportions with confidence intervals for paper data. }
     \label{fig:fiber_lags_proportions}
\end{figure}

To transform the image to local directions and subsequently assess uniform distribution of those directions, we maximize the norm-difference of observations at each point $x$ and its neighbors $\mathbb H^{(x)}$. The lags over which the maximization is performed are chosen to be sufficiently long to reliably leave the fiber when they point orthogonally to its direction in the $x$-$y$ plane or when the fiber is not purely horizontal, while remaining short enough to avoid frequently picking up neighboring fibers. Beyond their length, we may also choose their direction. We select one set of lags that measures anisotropy close to the cardinal directions ($\mathbb H_1$), one that measures close to the diagonals ($\mathbb H_3$), and one that is in between ($\mathbb H_2$):
\begin{align*}
&\mathbb H_1 \coloneqq \{(\pm 2, \pm8), (\pm8, \pm2)\}, && \mathbb H_2 \coloneqq \{(\pm 5, \pm10), (\pm10, \pm5)\}, \\
&\mathbb H_3 \coloneqq \{(\pm 8, \pm 11), (\pm 11, \pm 8)\}.
\end{align*}
Figure~\ref{fig:fiber_lags} illustrates these sets by overlaying them on a slice of a stochastic fiber model realization. Furthermore, we select a square sub-domain  $D$ of size $16 \times 16$. The empirical size under the null hypothesis remained fairly stable for window sizes varying between $10\times 10$ and $50 \times 50$.

\begin{figure}[!htb]
    \centering
    \subfloat[Alternative hypothesis: Fibers run mostly horizontally, but concentrate in a single direction. The concentration increases with $c$. At $c = 0$, the field is planar isotropic. \label{fig:fiber_power_c}]{%
        \begin{minipage}[b]{0.47\textwidth}
            \centering
            \includegraphics[width=\textwidth]{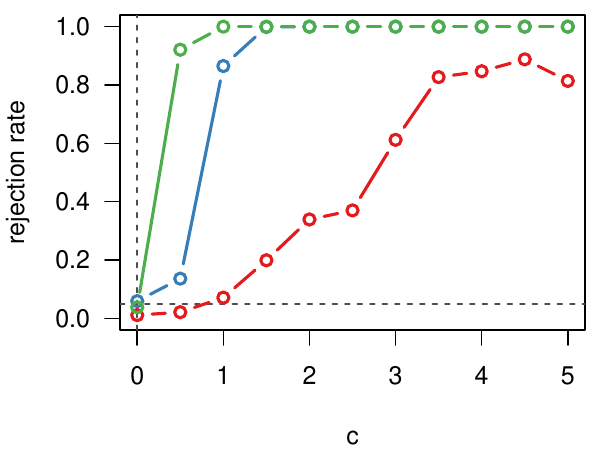}
        \end{minipage}%
    }
    \hfill
    \subfloat[Alternative hypothesis: Fibers run mostly horizontally in $y$-direction, but have slope $\theta$ in $x$-direction. At $\theta = 0$, the field is planar isotropic. \label{fig:fiber_power_angle}]{%
        \begin{minipage}[b]{0.47\textwidth}
            \centering
            \includegraphics[width=\textwidth]{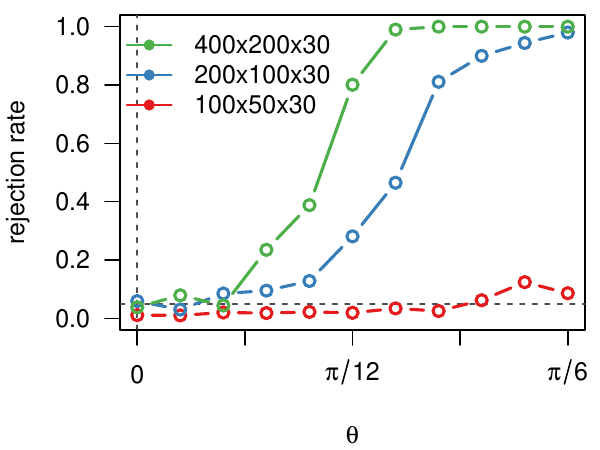}
        \end{minipage}%
    }
    \caption{Empirical power and size of UNI applied to the simulated paper material based on 1000 repetitions at significance level $\alpha = 0.05$ (marked with a horizontal line in each plot). Kinks in the lines may be further explained by instabilities in the estimation of $\lambda(\Xi_0)$.}
     \label{fig:fiber_power}
    
\end{figure}

Figure~\ref{fig:fiber_power} depicts empirical size of the test, as well as power under both alternatives. Clearly, increasing sample size improves power. Furthermore, the alternative of tilting the circular region of concentration by increasing $\theta$ is harder to detect and the test exhibits low power for  small sample size. 

We even observe just negligible power up to a certain angle $\theta$. Suppose that the measurement of the paper sample is not perfectly aligned with the $xy$-plane, but instead has a certain tilt, then its interior (with respect to $z$-direction) structure is related to that of a synthetically generated dataset with $\theta$ equal to that angle of tilt. Hence, the result indicates that the testing procedure exhibits no strong sensitivity with respect to such tilts.

\subsection{Evaluation of isotropy on the real data}
Finally, we apply the testing procedure to the dataset outlined in Section~\ref{sec:real_data}. To exclude influences caused of edge effects, the image is cropped in $z$-direction to consider only the $60$ interior pixels used to tune the stochastic fiber model. To be consistent with the choice of hyperparameters in Section~\ref{se:stoch_fiber_model}, we multiply all previously used lag sets and subwindow dimensions by~$5$. 

Recall that our test requires a stationary field. Stationarity in the $xy$-plane is well expected, see e.g.\  Section~3.1 of \cite{neumann2021variability}.
Additionally, we test this assumption layer-by-layer (in $z$-direction) using a stationarity test for textured images by \cite{taylor_test_2014}. The test is implemented in the {\tt R}~package \texttt{LS2Wstat} \citep{taylor_ls2wstat_2023}.  The outcome confirms that there is no signs for a violation of stationarity. 

After non-parallelized execution time of approximately 10 seconds on a standard laptop (eleventh generation Intel i5-1135G7 CPU), the $p$-value output is numerically $0$ (\texttt{< 2.2e-16}). Hence, our results indicate that there is strong evidence of anisotropy present in the dataset. This is in line with considering commercially manufactured unbleached paper sheets subjected to hard-nip calendering~\citep{neumann.2021}. For commercially manufactured paper, anisotropic effects naturally arise during the production process~\citep{niskanen1989}.

\begin{figure}[!htb]
    \centering
    \includegraphics[width=\linewidth]{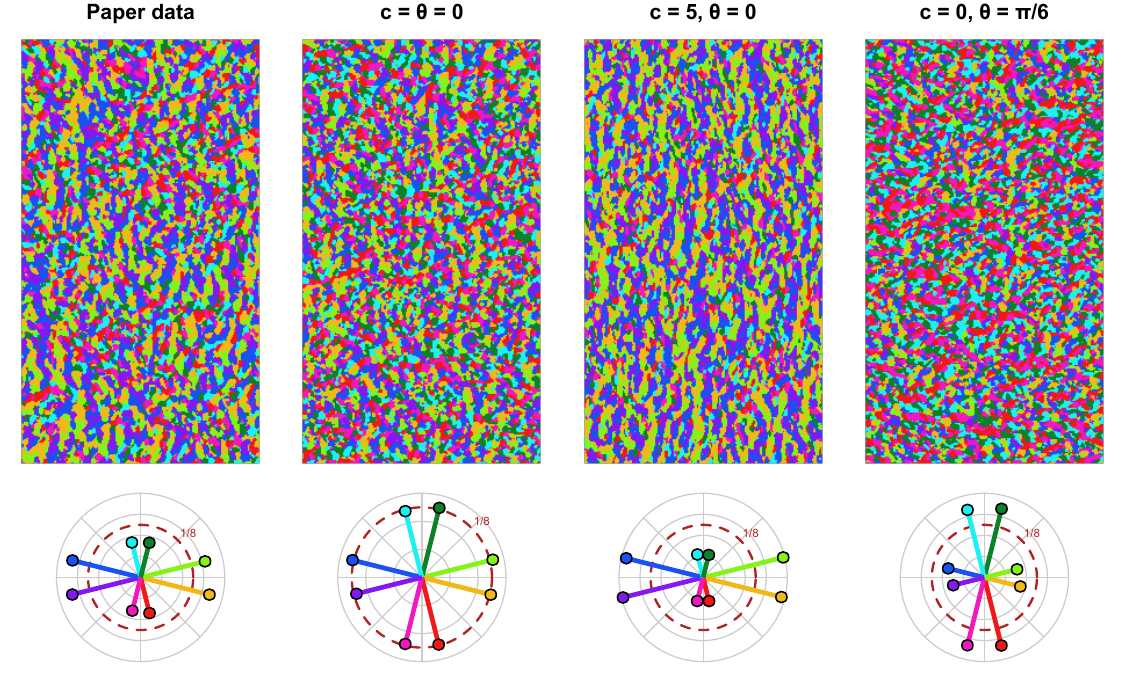}
    \caption{Top row: Maximizing directions selected from $\mathbb H_1$ for the paper-microstructure dataset as well as exemplary realizations of the fiber model under the null hypothesis ($c = \theta = 0$) and the two extremal alternatives ($c = 5$ and $\theta = \pi/6$, respectively). Bottom row: Legend indicating how colors used in top row correspond to elements of $\mathbb H_1$. The direction of the radial lines corresponds to that of the lag vector. The length of radial lines is given by the selection proportions. The red line is the length under uniform proportions (1/8).}
     \label{fig:fiber_argmax_image}    
\end{figure}

Figure~\ref{fig:fiber_lag_selection_proportions} further illustrates the respective proportions of how often each lag appears as a maximizer. Additionally, a joint hyperrectangular $95\%$ asymptotic confidence region for the proportions is computed based on the asymptotic Gaussian distribution of $S$ and visualized using individual confidence intervals for each proportion. These are found using the R package \texttt{mvtnorm}~\citep{package_mvtnorm}. Note that they are thus not necessarily a superset of the retention region of the proposed hypothesis test.
According to Figure~\ref{fig:fiber_lag_selection_proportions}, it appears that the violation is more in shorter lag lengths ($\mathbb H_1$) or along the cardinal directions rather than longer lag lengths or along the diagonals. Figure~\ref{fig:fiber_argmax_image} illustrates the process $M_0(x)$, defined through the lags $\mathbb H_1$ on $\mathcal D_0$, and compares it with our stochastic fiber models under three different parameterizations. The empirical data exhibit a pattern most similar to that of the fiber model with parameters $c=5$ and $\theta=0$, which corresponds to a tendency for the fibers to be oriented vertically (parallel to the $y$-axis).

\section{Discussion and future work} \label{sec:discussion}

In the present paper, we developed a hypothesis test for assessing the isotropy of abstract-valued random fields, with a particular focus on function-valued fields. We applied this methodology to evaluate the planar isotropy of microstructures in paper-based materials. Notably, the test applies to random fields in general spaces equipped with a semi-metric. Consequently, this approach provides a tool to justify simplifying assumptions regarding the spatial dependence structure for a very general class of fields.

While similar tests exist for spatio-temporal data, our setting differs fundamentally: we consider the temporal axis to be a fixed, non-expanding interval. Therefore, we do not require stationarity in time.

Furthermore, we provide graphical diagnostic tools in the form of simultaneous confidence bands. If the test rejects the null hypothesis of isotropy, these bands help identify which spatial lags are selected significantly more or less often than expected under a uniform distribution. %

The finite-sample performance of the test was investigated in both scalar and function-valued (specifically, high-dimensional multivariate) settings. In the scalar case, the proposed test exhibits computation times and empirical power comparable to the SIS test \citep{weller_nonparametric_2020}. While the GSC test \citep{guan_nonparametric_2004} is generally more powerful, it requires substantially higher computation times. For large sample sizes, both our methodology and the SIS test achieve a power comparable to the GSC test.

The implementation of the test is publicly available in the R package \texttt{unisono} \citep{baumhakel_baumhakelunisono_2026}, which is currently being prepared for release on CRAN.

To support our methodology, we provide a theoretical framework in the appendix. In this paper, our approach is restricted to the gridded observation setting. A natural direction for future theoretical work is extending this methodology to fields observed on a continuum or on random locations. Additional methodological improvements include implementing edge corrections %
and developing more sophisticated procedures for handling missing data in the interior. Currently, this is only addressed implicitly by treating missing observations as part of the domain boundary. Finally, it may be beneficial to explore self-normalization techniques to bypass the estimation of the long-run covariance, see e.g., \citet{zhang_selfnormalization_2014}. 

\section*{Declarations}

\paragraph{Declaration of generative AI and AI-assisted technologies in the writing process.}
 During the preparation of this work, the authors used ChatGPT and Claude to assist with mathematical exploration, language editing, and consistency checks. The authors independently verified all mathematical claims, derivations, and proofs and reviewed and edited all AI-assisted text. The authors take full responsibility for the content of the article.

\paragraph{Acknowledgements.} The authors thank Karin Zojer and Ingo Manke for providing 3D image data of a compressed paper sheet. The computational results have been achieved using the Austrian Scientific Computing (ASC) infrastructure.

\printbibliography

\newpage

\section{Appendix}
\input{proofs}

\end{document}

%% file: figures/grid_domains.tex
\begin{tikzpicture}[
    gridpoint/.style={circle, draw=black, inner sep=2pt, line width=1pt},
    gridpointthick/.style={circle, fill=black, inner sep=5pt},
    innerPoly/.style={draw=black, line width=3pt},
    outerPoly/.style={draw=black!55, line width=3pt, line join=round},
    axisLine/.style={-{Stealth[scale=1.5]}, line width=2pt, black},
    dpoint/.style={circle, fill=black!55, inner sep=2.7pt}, %
    endpoint/.style={
        circle, 
        draw=black, 
        fill=white, %
        inner sep=4pt,
        line width=1.5pt
    },
    d0point/.style={circle, fill=black, inner sep=2.7pt} %
]
    \clip (-6.5, -2.5) -- (-6.5, 12) -- (12, 12) -- (12, -2.5) -- cycle;

    \draw[axisLine] (-7.5, 0) -- (11, 0) node[below, font=\huge, text=black] {$x_1$};
    \draw[axisLine] (0, -7.5) -- (0, 11) node[left, font=\huge, text=black] {$x_2$};

    \foreach \x in {-6,-5,...,10} {
        \foreach \y in {-6,-5,...,10} {
            \node[gridpoint] at (\x, \y) {};
        }
    }

    \coordinate (A) at (-3.789, 2.728);
    \coordinate (B) at (-0.199, 7.754);
    \coordinate (C) at (7.776, 6.355);
    \coordinate (D) at (7.065, -1.045);
    \coordinate (E) at (3.534, -3.869);
    \coordinate (F) at (-2.386, -3.164);

    \draw[innerPoly] (A) -- (B) -- (C) -- (D) -- (E) -- (F) -- cycle;

    \coordinate (A_out) at (-6.2, 3.2);
    \coordinate (B_out) at (-1.2, 10.2);
    \coordinate (C_out) at (10.2, 8.2);
    \coordinate (D_out) at (9.2, -2.2);
    \coordinate (E_out) at (4.2, -6.2);
    \coordinate (F_out) at (-4.2, -5.2);

    \draw[outerPoly] (A_out) -- (B_out) -- (C_out) -- (D_out) -- (E_out) -- (F_out) -- cycle;

    \begin{scope}
        \clip (A_out) -- (B_out) -- (C_out) -- (D_out) -- (E_out) -- (F_out) -- cycle;
        
        \foreach \x in {-10,-9,...,10} {
            \foreach \y in {-10,-9,...,10} {
                \node[dpoint] at (\x, \y) {};
            }
        }
    \end{scope}

    \begin{scope}
        \clip (A) -- (B) -- (C) -- (D) -- (E) -- (F) -- cycle;
        
        \foreach \x in {-10,-9,...,10} {
            \foreach \y in {-10,-9,...,10} {
                \node[d0point] at (\x, \y) {};
            }
        }
    \end{scope}

        \draw (-2.6, 7.4) node[
            text=black!55, 
            font=\huge, 
            fill=white,          %
            draw=black!55,        %
            line width=2pt,    %
            inner sep=10pt,       %
            rounded corners=4pt  %
        ] {$\mathcal{D}$};

        \draw (-1, 5.4) node[
            text=black, 
            font=\huge, 
            fill=white,          %
            draw=black,        %
            line width=2pt,    %
            inner sep=10pt,       %
            rounded corners=4pt  %
        ] {$\mathcal{D}_0$};

    \coordinate (X) at (0,0);
    \coordinate (Y) at (4,6);
    \coordinate (Z) at (6,2);

    \foreach \origin in {(X), (Y), (Z)} {
         \draw[line width=2pt, draw=black!50, dashed] \origin circle (2.2360);
    
        \foreach \v in {(1,2), (2,1), (-1,2), (-2,1), (-1,-2), (-2,-1), (1,-2), (2,-1)} {
            \draw[-{Stealth[scale=1.3]}, 
                  line width=1.5pt, 
                  draw=Black!50,
                  shorten > = 5pt,
                  preaction={draw=white, line width=8pt, -}] %
                  \origin -- ++ \v 
                  node[endpoint] {} node[d0point] {};
        }
        \node[gridpointthick] at \origin {};
    }

    \node[dpoint] at (3,8) {};
    \node[dpoint] at (5,8) {};
    \node[dpoint] at (6,7) {};
    
    \node[dpoint] at (8,3) {};
    \node[dpoint] at (8,1) {};

\end{tikzpicture}

%% file: figures/grid_domains_2.tex
\begin{tikzpicture}[
    gridpoint/.style={circle, draw=black, inner sep=2pt, line width=1pt},
    gridpointthick/.style={circle, fill=black, inner sep=5pt},
    innerPoly/.style={draw=black, line width=3pt},
    outerPoly/.style={draw=black!55, line width=3pt, line join=round},
    axisLine/.style={-{Stealth[scale=1.5]}, line width=2pt, black},
    dpoint/.style={circle, fill=black!55, inner sep=2.7pt}, %
    endpoint/.style={
        circle, 
        draw=black, 
        fill=white, %
        inner sep=4pt,
        line width=1.5pt
    },
    d0point/.style={circle, fill=black, inner sep=2.7pt} %
]
    \clip (-6.5, -2.5) -- (-6.5, 12) -- (12, 12) -- (12, -2.5) -- cycle;

    \draw[axisLine] (-7.5, 0) -- (11, 0) node[below, font=\huge, text=black] {$x_1$};
    \draw[axisLine] (0, -7.5) -- (0, 11) node[left, font=\huge, text=black] {$x_2$};

    \foreach \x in {-6,-5,...,10} {
        \foreach \y in {-6,-5,...,10} {
            \node[gridpoint] at (\x, \y) {};
        }
    }

    \coordinate (A) at (-4.675, 2.902);
    \coordinate (B) at (-0.567, 8.653);
    \coordinate (C) at (8.667, 7.033);
    \coordinate (D) at (7.850, -1.469);
    \coordinate (E) at (3.779, -4.726);
    \coordinate (F) at (-3.053, -3.912);

    \draw[innerPoly] (A) -- (B) -- (C) -- (D) -- (E) -- (F) -- cycle;

    \coordinate (A_out) at (-6.2, 3.2);
    \coordinate (B_out) at (-1.2, 10.2);
    \coordinate (C_out) at (10.2, 8.2);
    \coordinate (D_out) at (9.2, -2.2);
    \coordinate (E_out) at (4.2, -6.2);
    \coordinate (F_out) at (-4.2, -5.2);

    \draw[outerPoly] (A_out) -- (B_out) -- (C_out) -- (D_out) -- (E_out) -- (F_out) -- cycle;

    \begin{scope}
        \clip (A_out) -- (B_out) -- (C_out) -- (D_out) -- (E_out) -- (F_out) -- cycle;
        
        \foreach \x in {-10,-9,...,10} {
            \foreach \y in {-10,-9,...,10} {
                \node[dpoint] at (\x, \y) {};
            }
        }
    \end{scope}

    \begin{scope}
        \clip (A) -- (B) -- (C) -- (D) -- (E) -- (F) -- cycle;
        
        \foreach \x in {-10,-9,...,10} {
            \foreach \y in {-10,-9,...,10} {
                \node[d0point] at (\x, \y) {};
            }
        }
    \end{scope}

        \draw (-2.6, 7.4) node[
            text=black!55, 
            font=\huge, 
            fill=white,          %
            draw=black!55,        %
            line width=2pt,    %
            inner sep=10pt,       %
            rounded corners=4pt  %
        ] {$\mathcal{D}$};

        \draw (-2, 5.4) node[
            text=black, 
            font=\huge, 
            fill=white,          %
            draw=black,        %
            line width=2pt,    %
            inner sep=10pt,       %
            rounded corners=4pt  %
        ] {$\mathcal{D}_0$};

    \coordinate (X) at (0,0);
    \coordinate (Y) at (4,6);
    \coordinate (Z) at (6,2);
    \coordinate (Z1) at (8,7);
     
    \foreach \origin in {(X), (Y), (Z), (Z1)} {
         \draw[line width=2pt, draw=black!50, dashed] \origin circle (1.414213);
        \foreach \v in {(1,1), (1,-1), (-1,1), (-1,-1)} {
            \draw[-{Stealth[scale=1.3]}, 
                  line width=1.5pt, 
                  draw=Black!50,
                  shorten > = 5pt,
                  preaction={draw=white, line width=8pt, -}] %
                  \origin -- ++ \v 
                  node[endpoint] {} node[d0point] {};
        }
        \node[gridpointthick] at \origin {};
    }

    \node[dpoint] at (7,8) {};
    \node[dpoint] at (9,8) {};
    \node[dpoint] at (9,6) {};

\end{tikzpicture}

%% file: proofs.tex
\subsection{Necessary properties for $\mathbb H^{(0)}$} \label{sec:lag_necess_cond}
The following lemmas give necessary conditions for $\mathbb H^{(0)}$ to be
transitive under its orthogonal symmetry group. 

\begin{lemma}[Equal norms and finiteness]\label{lem:equal-norm}
Let $\mathbb H\subset\mathbb Z^d$ be transitive under its orthogonal symmetry
group. Then there exists $r\geq 0$ such that $\|h\|=r$ for all
$h\in\mathbb H$, i.e.\ $\mathbb H\subseteq\mathbb S^{d-1}_0(r)\cap\mathbb Z^d$. In particular, $\mathbb H$ is finite.
\end{lemma}

\begin{proof}
Let $h,\tilde h\in\mathbb H$. By transitivity there exists $Q\in O(d)$ with
$Q\mathbb H=\mathbb H$ and $Qh=\tilde h$. Hence $\|\tilde h\|=\|Qh\|=\|h\|$. Thus
$\mathbb H$ is a subset of the sphere $\mathbb S^{d-1}_0(r)$, which is bounded
and therefore contains only finitely many points of $\mathbb Z^d$.
\end{proof}

\begin{lemma}[Equal spherical Voronoi measures]\label{lem:voronoi}
Let $d\geq 2$ and let $\mathbb H=\{h_1,\ldots,h_m\}\subset\mathbb Z^d$ be
transitive under its orthogonal symmetry group, with common norm $r>0$
(Lemma~\ref{lem:equal-norm}). For $i=1,\ldots,m$ define the spherical
Voronoi cell
\[
\mathcal C_i=\bigl\{x\in\mathbb S^{d-1}_0(r):\ \|x-h_i\|\leq\|x-h_j\|
\ \text{for all } j=1,\ldots,m\bigr\},
\]
and let $\nu$ denote the surface measure on $\mathbb S^{d-1}_0(r)$. Then
\[
\nu(\mathcal C_i)=\frac{\nu\bigl(\mathbb S^{d-1}_0(r)\bigr)}{m},
\qquad i=1,\ldots,m.
\]
\end{lemma}

\begin{proof}
Let $i,j\in\{1,\ldots,m\}$. By transitivity there exists $Q\in O(d)$ with
$Q\mathbb H=\mathbb H$ and $Qh_i=h_j$. Since $Q$ preserves distances and
permutes $\mathbb H$, it follows that $Q\mathcal C_i=\mathcal C_j$. As $\nu$ is
invariant under $O(d)$, $\nu(\mathcal C_i)=\nu(\mathcal C_j)$. Since all $h_i$
have norm $r$, the intersection $\mathcal C_i\cap\mathcal C_j$, $i\neq j$,
lies in the hyperplane $\{x:\langle x,h_i-h_j\rangle=0\}$ and is thus a
$\nu$-null set. Hence the cells partition $\mathbb S^{d-1}_0(r)$ up to a
$\nu$-null set, which proves the claim.
\end{proof}

\begin{lemma}[Identical distance profiles]\label{lem:distance-profile}
Let $\mathbb H=\{h_1,\ldots,h_m\}\subset\mathbb Z^d$ be transitive under its
orthogonal symmetry group. Then the multisets
\[
\{\!\{\|h_i-h_j\|:\ j=1,\ldots,m\}\!\} \quad \text{and} \quad \{\!\{\langle h_i,h_j\rangle:\ j=1,\ldots,m\}\!\}
\]
do not depend on $i$.
\end{lemma}

\begin{proof}
Let $i,i'\in\{1,\ldots,m\}$ and choose $Q\in O(d)$ with $Q\mathbb H=\mathbb H$
and $Qh_i=h_{i'}$. Then $h_j\mapsto Qh_j$ is a permutation of $\mathbb H$, and
\[
\|h_{i'}-Qh_j\|=\|Q(h_i-h_j)\|=\|h_i-h_j\|,
\qquad j=1,\ldots,m.
\]
Hence the multisets of norms coincide for $i$ and $i'$. The same holds true for the scalar products, since by
Lemma~\ref{lem:equal-norm},
$\|h_i-h_j\|^2=2r^2-2\langle h_i,h_j\rangle$.
\end{proof}

\subsection{Properties of $M_0(x)$}
\label{se:counterexample}

 \begin{proof}[Proof of Lemma~\ref{lem:uniformDistributionDiscrete}]
    Let $Q \in  O(d)$ and let $\mathbb H^{(0)} = \{h_1, \ldots, h_m\}$. Due to motion invariance, we obtain
    \[
        \big(\mathbbl d(\xi(h_i), \xi(0)) 
    \big)_{i = 1}^{m}
    \stackrel{\mathcal L}{=}
    \big(\mathbbl d(\xi(Qh_i), \xi(0))
    \big)_{i = 1}^{m},
    \]
    for $Q \in O(d)$ so that $\mathbb H^{(0)}$ is invariant under $Q$. Note that this invariance property is also valid for $Q'$. We may conclude %
    \begin{align*}
    M_0(x) = \argmax_{h \in \mathbb H^{(0)}}\mathbbl d(\xi(x+h), \xi(x)) &= Q\argmax_{h \in \mathbb H^{(0)}} \mathbbl d(\xi(x+Qh), \xi(x)) \\
    &\stackrel{\mathcal L}{=} Q~\argmax_{h \in \mathbb H^{(0)}} ~\mathbbl d(\xi(x+h), \xi(x))
   = Q M_0(x).
   \end{align*}
   Take any $h,\tilde h\in \mathbb H^{(0)}$ and $Q$ such that $Q'h=\tilde h$ and $\mathbb H^{(0)}$ is invariant under $Q$, which exists by assumption.
     Then,
    \[
    \prob{M_0(x) = h} = \prob{QM_0(x) = h} = \prob{M_0(x) = \tilde h}.
    \]
\end{proof}

The following example illustrates, why we require $\mathbb{H}^{(0)}$ to be transitive under its orthogonal symmetry group. Subsequently we denote by $e_i$ the $i$-th canonical basis vector.

\begin{example}\label{ex:nonuniform-gaussian}
Let $Z$ be a stationary Gaussian random field on $\mathbb Z^2$ with
$\cov(Z(x),Z(y))=\psi_Z(\|x-y\|)$, where $\psi_Z(0)=1$ and
$\rho\coloneqq\psi_Z(1)\in(0,1)$. Let $Z^{00},Z^{01},Z^{10},Z^{11}$ be
independent copies of $Z$ and define
\[
\xi(x)\coloneqq Z^{\,x\bmod 2}\bigl(\lfloor x/2\rfloor\bigr),
\qquad x\in\mathbb Z^2,
\]
where $\bmod$ and $\lfloor\cdot\rfloor$ act componentwise. Let
$\mathbb H^{(0)}\coloneqq\{e_1,-e_1,e_2\}$, which is not transitive under its
orthogonal symmetry group. Then $\xi$ is motion invariant, but
$\prob{M_0(x)=e_2}>1/3$, so $M_0(x)$ is not uniformly distributed on
$\mathbb H^{(0)}$.
\end{example}

\begin{proof}
We first discuss motion invariance. It
suffices to show that $\cov(\xi(x),\xi(y))$ depends only on $\|x-y\|$. If
$x\not\equiv y\pmod 2$, then $\xi(x)$ and $\xi(y)$ stem from independent
copies, and the covariance is zero. If $x\equiv y\pmod 2$, then
$\lfloor x/2\rfloor-\lfloor y/2\rfloor=(x-y)/2$, and the covariance is
$\psi_Z(\|x-y\|/2)$. Since squares of odd integers are $\equiv1$ and squares
of even integers are $\equiv0\pmod 4$, the case $x\equiv y\pmod 2$ occurs if
and only if $\|x-y\|^2\equiv0\pmod 4$. Hence the covariance depends only on
$\|x-y\|$.

$\mathbb H^{(0)}$ is not transitive under its orthogonal symmetry group due to Lemma~\ref{lem:voronoi}. We subsequently show that $M_0$ is not uniformly distributed. By stationarity, we may take $x=0$. Assume without loss of generality that $Z$ is centered. The four
relevant values  are
\[
\xi(0)=Z^{00}(0),\quad \xi(e_2)=Z^{01}(0),\quad
\xi(e_1)=Z^{10}(0),\quad \xi(-e_1)=Z^{10}(-e_1).
\]
All four are standard normal and independent, except for the pair
$\bigl(\xi(e_1),\xi(-e_1)\bigr)$, which has correlation $\rho$. Ties occur
with probability zero, so $M_0(0)=e_2$ holds if and only if both
$\xi(e_1)$ and $\xi(-e_1)$ lie in the random interval
\[
\mathcal J\coloneqq\bigl(\xi(0)-R,\ \xi(0)+R\bigr),\qquad R\coloneqq|\xi(e_2)-\xi(0)|.
\]
The interval $\mathcal J$ is independent of $\bigl(\xi(e_1),\xi(-e_1)\bigr)$, and
$R>0$ almost surely. Hence
\[
\prob{M_0(0)=e_2}=\mathbb E\bigl[p_\rho(\mathcal J)\bigr],\qquad
p_\rho(J)\coloneqq\prob{\zeta_1\in J,\ \zeta_2\in J},
\]
where $(\zeta_1,\zeta_2)$ is jointly normal with correlation
$\rho$ and has  standard normal marginals. Note that the law of $\mathcal J$ does not depend on $\rho$.

For $\rho=0$, the four values are i.i.d., so by symmetry each of the three
lags is the maximizer with probability $1/3$. It therefore suffices to show
that $p_\rho(J)$ is strictly increasing in $\rho\in[0,1)$ for every bounded
interval $J$ of positive length. By Mehler's formula
\citep[Theorem 2.4.1]{bryc_normal_1995},
\[
p_\rho(J)=\sum_{n=0}^\infty\frac{\rho^n}{n!}\,c_n(J)^2,
\qquad c_n(J)\coloneqq\int_J \mathrm{He}_n(s)\varphi(s)\,ds,
\]
where $\varphi$ is the standard normal density and $\mathrm{He}_n$ are the
Hermite polynomials. All coefficients are nonnegative. Moreover, $c_n(J)\neq0$
for some $n\geq1$, since otherwise $\mathbbm 1_J$ would be orthogonal to all
nonconstant Hermite polynomials and thus almost everywhere constant. Hence
$p_\rho(J)$ is strictly increasing in $\rho\in[0,1)$, and
$\prob{M_0(0)=e_2}>1/3$ for $\rho\in(0,1)$.
\end{proof}

\subsection{Relation between $H_0$ and $H_0'$} \label{sec:appendix_h0_relation}

The following results support the claims in Example~\ref{ex:animult}.

\begin{proposition}\label{prop:animult}
Consider the setup described in Example~\ref{ex:animult}. Then, for all $x\in\mathbb Z^d$,
\[
\operatorname{sign}\Bigl(\prob{M_0(x)=h_1}-\tfrac12\Bigr)
=\operatorname{sign}\bigl(h_1'Ah_1-h_2'Ah_2\bigr).
\]
\end{proposition}

\begin{proof}

Since $\Theta$ is orthogonal,
we have for any $u,v\in\mathbb{Z}^d$ that $\mathbbl{d}(\Theta'\xi(u),\Theta'\xi(v))=\mathbbl{d}(\xi(u),\xi(v))$ when $\mathbbl{d}$ is the Euclidean distance. We
may therefore assume $\Theta=I$, so that the components $\xi_1,\ldots,\xi_p$
are independent and univariate fields as in Example~\ref{ex:ani}, with covariance
functions $\psi_j(h'Ah)$. By stationarity it suffices to consider $x=0$.  

Define $U_j$ and $V_j$ as in Example~\ref{ex:ani},
with  $\xi$ replaced by $\xi_j$. Then
\[
Y\coloneqq\|\xi(h_1)-\xi(0)\|^2-\|\xi(h_2)-\xi(0)\|^2=\sum_{j=1}^pU_jV_j,
\qquad \prob{M_0(0)=h_1}=\prob{Y>0},
\]
where the pairs $(U_j,V_j)$ are independent and
$\cov(U_j,V_j)=2\bigl(\psi_j(h_2'Ah_2)-\psi_j(h_1'Ah_1)\bigr)$. Since every
$\psi_j$ is strictly decreasing, all these covariances share the sign of
$h_1'Ah_1-h_2'Ah_2$.

Next, we write $U_j V_j$ as a linear combination of $\chi_1^2$ random variables. To this end, denote $Z_{j,+} \coloneqq U_j/\sqrt{\var(U_j)}+  V_j/\sqrt{\var(V_j)}$ and $Z_{j,-} \coloneqq U_j/\sqrt{\var(U_j)}-  V_j/\sqrt{\var(V_j)}$, which are independent, centered Gaussian variables. Then it can be easily verified that
$
U_jV_j = \frac14 \sqrt{\var(U_j) \var(V_j)} (Z_{j,+}^2 - Z_{j,-}^2),
$
and consequently there exist constants $a_j,b_j>0$ so that we may write
$$
  Y=\sum_{j=1}^p\bigl(a_j\zeta^2_j-b_j \tilde\zeta_j^2\bigr),
  $$
  where $\zeta^2_1,\tilde \zeta_1^2,\ldots,\zeta_p^2,\tilde \zeta_p^2\stackrel{\text{iid}}{\sim}\chi^2_1$. Moreover, it can readily verified that
  $a_j-b_j= \frac12 \cov(U_j,V_j)$.

Suppose now $h_1'Ah_1\geq h_2'Ah_2$, so that $a_j\geq b_j$ for all $j$. Set
$Y^-\coloneqq\sum_j(b_j\zeta^2_j-a_j\tilde \zeta^2_j)$. Exchanging $\zeta^2_j$ and $\tilde \zeta^2_j$
shows $Y^-\overset{\mathcal L}{=}-Y$. Moreover,
$Y-Y^-=\sum_j(a_j-b_j)(\zeta^2_j+\tilde \zeta^2_j)\geq0$. Hence
\[
\prob{Y>0}-\prob{Y<0}=\prob{Y>0}-\prob{Y^->0}=\prob{Y^-\leq0<Y}\geq 0.
\]
If $h_1'Ah_1=h_2'Ah_2$, then $a_j=b_j$, so $Y=Y^-$ and $\prob{Y>0}=\frac12$.
If $h_1'Ah_1>h_2'Ah_2$, then at $\zeta^2_j=\tilde \zeta^2_j=1$ for all $j$ we have
$Y=\sum_j(a_j-b_j)>0>Y^-$. By continuity, this holds on an open
neighborhood, which has positive probability. Hence $\prob{Y>0}>\frac12$.
The case $h_1'Ah_1<h_2'Ah_2$ follows by exchanging $h_1$ and $h_2$.
\end{proof}

\begin{corollary}\label{cor:animult}
Let $\xi$ be as in Proposition~\ref{prop:animult} on $\mathbb Z^d$, $d\geq2$,
and for $1\leq i<j\leq d$ define
\[
\mathbb H^{(0)}_{j}\coloneqq\{e_1, e_j\},
\qquad
\widetilde{\mathbb H}^{(0)}_{ij}\coloneqq\{e_i+e_j,\ e_i-e_j\}.
\]
Each of these sets is transitive under its orthogonal symmetry group. If $A$
is not a positive multiple of the identity matrix, then $M_0(x)$ is not uniformly distributed with
respect to at least one of these $(d-1)(d+2)/2$ lag sets. Conversely, if $A$ is a positive multiple of the identity,
then $\xi$ is isotropic and $M_0(x)$ is uniform with respect to all of them.
Hence $H_0=H_0'$ if uniformity of $M_0(x)$ is required with respect to all
of these lag sets simultaneously.
\end{corollary}

\begin{proof}
As in Example~\ref{ex:lags} $\mathbb H^{(0)}$ and $\widetilde{\mathbb H}^{(0)}$ are transitive under their respective orthogonal symmetry groups. By Proposition~\ref{prop:animult}, $M_0(x)$ is uniform
on $\{h_1,h_2\}$ if and only if $h_1'Ah_1=h_2'Ah_2$. For the sets above,
\[
e_1'Ae_1-e_j'Ae_j=A_{1,1}-A_{j,j},
\qquad
(e_i+e_j)'A(e_i+e_j)-(e_i-e_j)'A(e_i-e_j)=4A_{i,j}.
\]
If $M_0(x)$ is uniform with respect to all of these sets, then all diagonal
entries of $A$ coincide and all off-diagonal entries vanish, i.e.\ $A=a \mathbb I$
with $a>0$ by positive definiteness, where $\mathbb I$ is the identity matrix. Conversely, if $A=a \mathbb I$, then $C(h)$ depends on $h$
only through $\|h\|$, so $\xi$ is isotropic and $M_0(x)$ is uniform by
$H_0\subset H_0'$.
\end{proof}

\subsection{Asymptotic justification under the null} \label{sec:appendix_asymptotic_null}

We assume a weak dependence structure for the field $\xi$ based on $\alpha$-mixing. To this end, suppose $I \subset \mathcal{D}$ and denote the $\sigma$-algebra generated by $\{\xi(x)\colon x\in I\}$ by $\mathcal{F}_I$. Then we require
$$
\alpha_\xi(u)\coloneqq \sup |\prob{E\cap E'}-\prob{E} \prob{E'}|\to 0 \quad (u\to \infty),
$$
at a certain rate, where the supremum is taken over all events $E\in \mathcal{F}_{I}$ and $E'\in \mathcal{F}_{I'}$ and all sets $I, I'\subset \mathcal{D}$, such that $\dist(I,I')\coloneqq \inf_{x\in I, y\in I'}\norm{x-y}\geq u$.  
Controlling dependence decay by an $\alpha$-mixing condition is common in isotropy testing and utilized by both \citet{guan_nonparametric_2004} and \citet{weller_nonparametric_2020}. Furthermore, $\alpha$-mixing is a comparably weak assumption within classical mixing conditions, see e.g.~\citet{bradley_introduction_2007}. If we assume that $\mathcal V = \mathbb R$ and $\xi$ is a Gaussian random field on $\mathbb Z^d,$ then $\cov(\xi(0), \xi(x)) = \mathcal O(|x|^{-(3d + \varepsilon)})$ implies the mixing condition $\alpha_\xi(u) = \mathcal O(u^{-(2d + \varepsilon)})$, see Section 2.1.1 in~\cite{doukhan.1994}.
 Furthermore, $m$-dependent fields fulfill the $\alpha$-mixing assumption (at any rate).

For asymptotic statements, the observation domain is denoted by $\mathcal{D}^k$ and is assumed to grow with  $k$. We assume that $\mathcal{D}^k$ is sufficiently regular so that for most $x\in \mathcal{D}^k$, the full set of neighbors $\mathbb{H}^{(x)}=x+\mathbb H^{(0)}$ is available. Analogously to $\mathcal D_0$, we denote the set of $x$ with this property by~$\mathcal{D}_0^k$. Formally we require the domain to grow regularly in the following sense \citep[p. 154]{nahapetian_limit_1991}:
\begin{assumption}\label{ass:D} The sets $(\mathcal{D}^k)_{k \geq 1}$ (thus also $(\mathcal{D}_0^k)_{k \geq 1}$) form a \textit{Van Hove sequence} in $\mathbb{Z}^d$.
\end{assumption}
Note that Assumption~\ref{ass:D} particularly ensures that $\lim_{k \rightarrow \infty} n(\mathcal D^k_0) / n(\mathcal D^k) = 1.$ For a detailed discussion of Van Hove sequences in $\mathbb R^d$ and $\mathbb Z^d,$ we refer to Chapter~3.1 in~\citet{bulinski.2007}. 

\begin{theorem}\label{thm:discreteUniformCLT}
    Let $\xi$ be a motion-invariant $\mathcal V$-valued random field on $\mathbb Z^d$ such that $\alpha_\xi(u)=\mathcal O(u^{-(2d+\varepsilon)})$ for some $\varepsilon >0$. Suppose that Assumption~\ref{ass:D} holds and that $\mathbb H^{(0)}$ is transitive under its orthogonal symmetry group. Then, 
    \begin{equation} \label{eq:latticeCaseTestCLT_k}
    S_k \coloneqq  \frac{1}{\sqrt{n({\mathcal{D}_0^k)}}}\sum_{x \in \mathcal{D}_0^k} \left[  \mathcal {I}(x) - \frac{1}{m} \ind_{m} \right],
    \end{equation}
    converges in law to a zero mean Gaussian vector,
    \[
    S_k \xrightarrow[k \to \infty]{\mathcal L} \mathcal N_m \left( 0, \Gamma \right), \quad \Gamma = \sum_{h \in \mathbb Z^d} \cov \left( \mathcal I(0), \mathcal I(h) \right).
    \]
    The matrix series defining $\Gamma$ is absolutely convergent. 
\end{theorem}

To show Theorem~\ref{thm:discreteUniformCLT}, we use a central limit theorem by \citet[Thm.~7.2.3]{nahapetian_limit_1991}. The following result is a special case of this theorem.
\begin{lemma}\label{lem:nahepitan_clt}
    Let $Y$ be a real-valued centered, stationary random field on $\mathbb Z^d$ such that for some $p > 2$, $\expv{\abs{Y(0)}^{p}} < \infty$ that fulfills an $\alpha$-mixing condition with $\alpha_Y(u) = \mathcal O(u^{-(\rho_d+\varepsilon)})$ for some $\varepsilon > 0$, where
    \[
    \rho_d = \max \left\{ \frac{p}{p-2}, 2d \right\}.
    \]
    Then for any sequence of sets $\mathcal D^k \subset \mathbb Z^d$ growing in the Van Hove sense, the central limit theorem
    \[
    n(\mathcal D^k)^{-\frac12} \sum_{h \in \mathcal D^k} Y(h) \xrightarrow[k \to \infty]{\mathcal L} \mathcal N(0,\sigma_Y^2),
    \]
    holds, where
    \[
    0 < \sigma_Y^2 = \sum_{h \in \mathbb Z^d} \expv{Y(0) Y(h)} < \infty.
    \]
\end{lemma}

\begin{proof}[Proof of Theorem~\ref{thm:discreteUniformCLT}]
    The mixing condition of $\xi$ transfers to $M_0(x)$ and thus also to the process $a'\mathcal I(x)$, for any $a\in\mathbb{R}^m$. In particular we obtain
    \[
    \alpha_{a' \mathcal I}(u+2r) \leq \alpha_{\xi}(u),
    \]
   for each $u \geq 1$. Furthermore, stationarity of $\xi$ also implies stationarity of $M_0$ and $a'\mathcal I$, which is an almost surely bounded random field, so the moment condition applies for arbitrary $p > 2$. By Lemma~\ref{lem:nahepitan_clt} we obtain that
    $
    a'S_k\xrightarrow[]{\mathcal L} \mathcal N(0, a' \Gamma a)
    $
    when $k\to\infty$. The multivariate CLT follows by the Cramér-Wold device.
\end{proof}

    \begin{corollary} \label{cor:TestWithEstimator_discrete}
    Suppose that $\widehat\Gamma_k$
    is a weakly consistent estimator of $\Gamma$ and that the assumptions of Theorem~\ref{thm:discreteUniformCLT}
    hold. Then \eqref{e:decisiond}
    is an asymptotically valid level-$\alpha$ test for $H_0'$.
    \end{corollary}

\subsection{Asymptotic justification under the alternative}~\label{s:consistency}
We establish consistency for
the weaker null hypothesis
\[
H_0'\colon M_0(x)\text{ is uniformly distributed}.
\]
 The asymptotic power under fixed and local alternatives is summarized in the following two propositions. The first is a consequence of the law of large numbers.
\begin{proposition}[Consistency against fixed alternatives]
\label{prop:fixed-alternative}
Suppose that $\xi$ is an anisotropic random field in $\mathbb Z^d$ that fulfills the remaining conditions of Theorem~\ref{thm:discreteUniformCLT}, with non-uniformly distributed $M_0$ in the sense that
\[
\Delta\coloneqq\mathbb E[\mathcal I(0)]-\frac1m\ind_m\neq 0.
\]
Let $\xi$ be observed on $\mathcal D^k$ according to Assumption~\ref{ass:D}. Then, the probability of rejecting $H_0'$ tends to $1$ as $k \to \infty.$
\end{proposition}
Next, we formulate a consistency result for local alternatives, i.e. settings, where anisotropy weakens with increasing sample size. 

\begin{proposition}[Power against local alternatives]
\label{prop:local-alternative}
Let $(\xi^{(k)})_{k \ge 1}$ be a sequence of stationary random fields where $\xi^{(k)}$ is observed on  $\mathcal D^k \subset \mathbb Z^d$ following Assumption~\ref{ass:D}. Let $\mathcal I^{(k)}$ be its associated index field. Suppose  that
\begin{equation} \label{eq:localAlternativeIndicator}
\mathbb E[\mathcal I^{(k)}(0)]-\frac1m\ind_m = n(\mathcal D^k)^{-1/2}\Delta_k, \quad \norm{\Delta_k} = 1,
\end{equation}
Let $\Gamma_k$ be the long-run covariance of $\mathcal I^{(k)}$ and assume  $\Gamma_k\to\Gamma_0$ as $k \to \infty$. Let further $\widehat \Gamma_k$ be weakly consistent estimators in the sense that $(\widehat \Gamma_k - \Gamma_k) \to 0$, in probability. 

Suppose $\sup_k \alpha_{\xi^{(k)}}(u) = \mathcal O(u^{-(2d+\varepsilon)})$ for some $\varepsilon > 0$.  Then, at significance level $\alpha$,
\[
\liminf_{k\to\infty}\mathbb P_{\xi^{(k)}}\bigl(H_0' \text{ is rejected }\bigr)>\alpha .
\]
\end{proposition}
\begin{proof}
By \eqref{eq:localAlternativeIndicator},
\[
\frac{1}{\sqrt{n(\mathcal D_0^k)}}\sum_{x\in\mathcal D_0^k}\left[\mathcal I^{(k)}(x)-\frac1m\ind_m\right]
=
\frac{1}{\sqrt{n(\mathcal D_0^k)}}\sum_{x\in\mathcal D_0^k}\left[\mathcal I^{(k)}(x)-\mathbb E\mathcal I^{(k)}(0)\right]+\sqrt{\frac{n(\mathcal D_0^k)}{n(\mathcal D^k)}}\,\Delta_k.
\]
The bounds $\sup_k\alpha_{\xi^{(k)}}(u)=\mathcal O(u^{-(2d+\varepsilon)})$ translate to the indicator fields and allow the triangular-array CLT of \citet{jenish_central_2009} to be applied to every scalar projection. Hence, for every $a\in\mathbb R^m$,
\[
a'\frac{1}{\sqrt{n(\mathcal D_0^k)}}\sum_{x\in\mathcal D_0^k}\left[\mathcal I^{(k)}(x)-\mathbb E\mathcal I^{(k)}(0)\right]
\xrightarrow{\mathcal L}\mathcal N\!\left(0,a'\Gamma_0a\right),
\]
since $\Gamma_k\to\Gamma_0$. Note that $n(\mathcal D_0^k)/n(\mathcal D^k)\to 1$ by Assumption~\ref{ass:D}. By the Cramér--Wold device, along every  subsequence $\Delta_{k_\ell}\to\Delta_\infty$,
\[
\frac{1}{\sqrt{n(\mathcal D_0^k)}}\sum_{x\in\mathcal D_0^k}\left[\mathcal I^{(k)}(x)-\frac1m\ind_m\right]
\xrightarrow{\mathcal L}\mathcal N_m(\Delta_\infty,\Gamma_0).
\]
Since $\|\Delta_k\|=1$, every subsequence admits a further subsequence converging to a limit with norm 1. Moreover, we have $\widehat\Gamma_k\xrightarrow{\mathbb P}\Gamma_0.$
Thus, by Slutsky's theorem, $T_k=\norm{S_k}^2$ converges along every such subsequence to a specific weighted non-central $\chi^2$-distribution, which stochastically strictly dominates the identically-weighted central $\chi^2$-distribution. This implies our claim.
\end{proof}

\subsection{Consistency of the long-run covariance estimator}

    To show  consistency, we require the following conditions for the basis sub-domains and shifts.

\begin{assumption} \label{ass:covEstimationShiftsDiscrete}
    For each $k\in\mathbb N$, let $D_k\subset\mathbb Z^d$ be a finite lattice set forming a Van Hove sequence, with $n(D_k)\to\infty$. Let $D_{k;\ell}=D_k+s_{k;\ell}\subset\mathcal D_0^k$, $\ell=1,\ldots,N_k$, be pairwise disjoint translates of $D_k$. Moreover, assume that $n(D_k)/N_k \to 0$ as $k \to \infty$.
\end{assumption}

\begin{theorem} \label{thm:discreteLongrunConsistent}
   Let $\xi$ be a stationary $\mathcal V$-valued random field on $\mathbb Z^d$ that fulfills a mixing condition with $\alpha_\xi(u)=\mathcal O(u^{-(d+\varepsilon)})$ for some $\varepsilon>0$. Let $((D_{k;\ell})_{\ell=1}^{N_k})_{k\geq1}$ be a sequence of families of subdomains for which Assumption~\ref{ass:covEstimationShiftsDiscrete} holds. Then the estimator  $\widehat{\Gamma}_k$ as described in Section~\ref{se:estcov} is weakly consistent.
\end{theorem}

\begin{proof}[Proof of Theorem~\ref{thm:discreteLongrunConsistent}]
    The estimator is positive semi-definite by the representation
    \[
    \widehat \Gamma_k=\frac{n(D_k)}{N_k-1}U_k'U_k,
    \qquad U_{k;\ell,i}\coloneqq \widehat\mu_{k;\ell,i}-\overline\mu_{k;i},
    \]
    for $i=1,\ldots,m$ and $\ell=1,\ldots,N_k$, where indices after the semicolon denote matrix and vector entries, respectively.

    To show consistency, we adapt the approach of \citet{pantle_estimation_2010} (specifically Lemma~2 and Theorem~3) to a random field observed on a lattice under a weak dependence condition. We establish entry-wise mean-square consistency by showing that the estimator is asymptotically unbiased and that the variances of its entries vanish as $k\to\infty$.

    Throughout the proof, let $\mathcal I_0(x)\coloneqq\mathcal I(x)- \expv{\mathcal I(0)}$ denote the centered index field. Since the estimator is invariant under a common deterministic shift of the block means, we may work with block means of $\mathcal I_0$. Write $n_k\coloneqq n(D_k)$ and $\overline D_k\coloneqq\bigcup_{\ell=1}^{N_k}D_{k;\ell}$, so that $n(\overline D_k)=N_kn_k$.

    Since $\mathcal I_0$ is a measurable transformation of $\xi$, it satisfies the same mixing-rate bound. In particular, $\sum_{h\in\mathbb Z^d}\alpha_{\mathcal I_0}(\norm h)<\infty$. Further, for every finite $D, D'\subset\mathbb Z^d$,
    \begin{equation} \label{eq:latticeSummability}
        \sum_{x\in D, y \in D'}\alpha_{\mathcal I_0}(\norm{x-y})
        \le n(D)\sum_{h\in\mathbb Z^d}\alpha_{\mathcal I_0}(\norm h).
    \end{equation}

    First, decompose the estimator into
    \begin{equation} \label{eq:gammaEstimatorDecomposition_discrete}
    \widehat \Gamma_k
    =\frac{n_k}{N_k}\sum_{\ell=1}^{N_k}\widehat\mu_{k;\ell}\widehat\mu_{k;\ell}'
    -\frac{n_k}{N_k(N_k-1)}\sum_{\ell\ne\ell'=1}^{N_k}
    \widehat\mu_{k;\ell}\widehat\mu_{k;\ell'}'
    =:S_{k,1}+S_{k,2}.
    \end{equation}
    By stationarity,
    \begin{align*}
        \Gamma-\expv{S_{k,1}}
        =\sum_{h\in\mathbb Z^d}\cov(\mathcal I_0(0),\mathcal I_0(h))
        \frac{n(D_k\setminus(D_k-h))}{n_k}.
    \end{align*}
    By \citet[Theorem~3.7]{bradley_introduction_2007}, the entries of $\cov(\mathcal I_0(0),\mathcal I_0(h))$ are absolutely bounded by $8\alpha_{\mathcal I_0}(\norm h)$ and are therefore absolutely summable. The Van Hove property and dominated convergence consequently yield $\expv{S_{k,1}}\to\Gamma$.

    For $S_{k,2}$, the mixing covariance inequality and \eqref{eq:latticeSummability} give, entry-wise,
    \begin{align*}
        \abs{\expv{S_{k,2;i,j}}}
        &\le \frac{8}{N_k(N_k-1)n_k}
        \sum_{\ell\ne\ell'=1}^{N_k}
        \sum_{\substack{x\in D_{k;\ell}\\y\in D_{k;\ell'}}}
        \alpha_{\mathcal I_0}(\norm{x-y}) \\
        &\le \frac{8}{N_k(N_k-1)n_k}
        \sum_{\ell = 1}^{N_k}
        \sum_{x\in D_{k;\ell}} \sum_{y \in \overline D_k}
        \alpha_{\mathcal I_0}(\norm{x-y}) \le \frac{8}{N_k-1}
        \sum_{h\in\mathbb Z^d}\alpha_{\mathcal I_0}(\norm h),
    \end{align*}
    which goes to $0$ as $k \to \infty$. 
    
    Next, we show that the variances vanish. For $S_{k,1}$, the mixing covariance inequality yields
    \[
    \var(S_{k,1;i,j})
    \le \frac{4}{N_k^2n_k^2}
    \sum_{\ell,\ell'=1}^{N_k}
    \sum_{\substack{x_1,y_1\in D_{k;\ell}\\x_2,y_2\in D_{k;\ell'}}}
    \alpha_{\mathcal I_0}(\{x_1,y_1\},\{x_2,y_2\})
    \norm{\mathcal I_0(0)}_\infty^4,
    \]
    where $\norm{\mathcal I_0(0)}_\infty^4\le1$. By monotonicity,
    \[
    \alpha_{\mathcal I_0}(\{x_1,y_1\},\{x_2,y_2\}) \le \sum_{\substack{x \in \{x_1, y_1\} \\ y \in \{x_2, y_2\}}} \alpha_{\mathcal I_0}(\norm{x-y}).
    \]
    Each summand on the right-hand side leaves at most $n_k^2$ choices for the unused indices. After collapsing the sum, by \eqref{eq:latticeSummability},
    \[
        \var(S_{k,1;i,j})
        \le \frac{16}{N_k^2}
        \sum_{x,y\in \overline D_k}\alpha_{\mathcal I_0}(\norm{x-y})
        \le \frac{16n_k}{N_k}
        \sum_{h\in\mathbb Z^d}\alpha_{\mathcal I_0}(\norm h)
        \to 0,
    \]
    by Assumption~\ref{ass:D}. Similarly, for $S_{k,2}$,
    \begin{align*}
        \var(S_{k,2;i,j})
        &\le \frac{16}{N_k^2(N_k-1)^2n_k^2}
        \sum_{x_1, x_2, y_1, y_2\in \overline D_k}\alpha_{\mathcal I_0}(\norm{x_1-x_2})\\
        &\le \frac{16N_kn_k}{(N_k-1)^2}
        \sum_{h\in\mathbb Z^d}\alpha_{\mathcal I_0}(\norm h)
        \to 0.
    \end{align*}
    Here, the first bound follows by expanding the covariance of the products, bounding the set-distance mixing coefficient by the sum of four pointwise coefficients as above, and dropping the restriction $\ell \neq \ell'$.

    We have thus shown that $\expv{S_{k,1}}\to\Gamma$, $\expv{S_{k,2}}\to0$, and the variances of all entries of $S_{k,1}$ and $S_{k,2}$ vanish. Consequently, $\widehat\Gamma_k$ converges entry-wise in mean square to $\Gamma$, and is therefore a weakly consistent estimator of $\Gamma$.
\end{proof}